\documentclass[twocolumn, twocolappendix]{aastex701}
\usepackage{amsmath}
\usepackage{subcaption}
\usepackage{placeins}

\begin{document}

\title{Cosmic Ray Transport in Protoplanetary Disk Winds: A Theoretical Framework}

\author[0009-0005-4398-0138]{Zisen Dai}
\affiliation{Zhili College, Tsinghua University, Beijing 100084, People's Republic of China}
\email{dzs23@mails.tsinghua.edu.cn}

\author[0000-0001-6906-9549]{Xue-Ning Bai}
\affiliation{Institute for Advanced Study, Tsinghua University, Beijing 100084, People's Republic of China}
\affiliation{Department of Astronomy, Tsinghua University, Beijing 100084, People's Republic of China}
\email{xbai@tsinghua.edu.cn}

\shorttitle{Cosmic Ray Transport in Disk Winds}
\shortauthors{Dai \& Bai}

\correspondingauthor{Zisen Dai}
\correspondingauthor{Xue-Ning Bai}

\begin{abstract}
Cosmic rays (CRs) are among the most important ionization sources in protoplanetary disks (PPDs), which set the coupling between gas and magnetic fields, with profound consequences for gas dynamics and angular momentum transport. However, how much of the interstellar CR flux reaches PPDs remains poorly known, especially because CRs must travel through a magnetized PPD wind. We follow test CR protons from an injection radius $R_{\rm inj}$ along one representative field line anchored at $R_0=1\,\mathrm{AU}$ in a magneto-thermal disk wind under the guiding-center approximation, and compute $\Phi$, the ratio of CR flux reaching the disk to the interstellar flux. We further incorporate pitch-angle scattering and cross-field diffusion, representing the unknown effect of turbulence with amplitude $\eta=(\delta B/B)^2$ under a prescribed spectrum.
In a laminar wind, while magnetic focusing and mirroring largely cancel, the CR flux can still be reduced by $\sim R_{\rm inj}/R_{\rm LC}$ due to energy loss through the wind, where $R_{\rm LC}$ is the field line's light cylinder radius.
In a turbulent wind, the CR flux is approximately set by the balance between wind advection and turbulent diffusion, with $\Phi\lesssim e^{-\mathcal{M}_R}$,
where $\mathcal{M}_R=\int|u_R|\,\mathrm{d}R/\kappa_{RR}$ is the radial modulation depth given the radial wind speed $u_R$ and projected radial diffusion coefficient $\kappa_{RR}$.
Toward strong turbulence, cross-field diffusion dominates and may retain the CR flux by reducing the CR path length (and $\mathcal{M}_R$) in the tightly wound helical field. In all regimes, the flux of lower-energy CRs is preferentially reduced/suppressed by the PPD wind, analogous to CR exclusion by the solar wind.
\end{abstract}


\section{Introduction}
\label{sec:intro}

The evolution of protoplanetary disks is regulated by the redistribution and removal of angular momentum, with profound implications for planet formation. It is well known that magnetic fields play a crucial role \citep{baiAngularMomentumTransport2026}. In sufficiently ionized gas, magnetohydrodynamic stresses can be generated by the magnetorotational instability (MRI), providing one major route to turbulent angular momentum transport \citep{balbusPowerfulLocalShear1991}. In much of the planet-forming region of protoplanetary disks, however, the gas is weakly ionized, which only partially couples to the magnetic field, described by three non-ideal MHD effects: Ohmic resistivity, ambipolar diffusion and the Hall effect \citep{wardleMagneticFieldsProtoplanetary2007,baiMAGNETOROTATIONALINSTABILITYDRIVENACCRETIONPROTOPLANETARY2011,baiHALLEFFECTCONTROLLEDGASDYNAMICS2014,lesurThanatologyProtoplanetaryDiscs2014}. Depending on the ionization level, these non-ideal effects can reduce or suppress MRI-driven turbulence \citep[e.g.,][]{baiEFFECTAMBIPOLARDIFFUSION2011}. Instead, it has been recognized that in most regions of protoplanetary disks (perhaps except for the innermost region $<1\,\mathrm{AU}$), angular momentum transport is primarily mediated by magnetized disk winds, as identified by a series of numerical studies \citep{baiWINDDRIVENACCRETIONPROTOPLANETARY2013,baiWINDDRIVENACCRETIONPROTOPLANETARY2013II,gresselGlobalSimulationsProtoplanetary2015,baiGlobalSimulationsInner2017,lesurSystematicDescriptionWinddriven2021,cuiGlobalThreedimensionalSimulations2021}.

Ionization in protoplanetary disks is supplied by several channels whose relative importance depends on location, column density, and stellar environment. Stellar ultraviolet photons can ionize the gas to a high degree of $\gtrsim10^{-4}$, but restricted to the disk atmosphere, up to $0.1$ g cm$^{-2}$ \citep{hollenbachDENSEPHOTODISSOCIATIONREGIONS1997,perez-beckerSURFACELAYERACCRETION2011}. Stellar X-rays can penetrate deeper (up to $\sim5$ g cm$^{-2}$ for the scattered component) but are still strongly attenuated by disk columns \citep{igeaXRayIonizationDisks1999,ercolanoXrayIonizationRates2013}. Short-lived radionuclides, mostly $^{26}$Al, provide a local ionization floor at a very low level of the order $10^{-19}$ s$^{-1}$ \citep[e.g.,][]{turnerENERGETICPROTONSRADIONUCLIDES2009}. Finally, external Galactic cosmic rays (CRs) and energetic particles are considered to contribute at the largest columns of up to $100$ g cm$^{-2}$ at the level of $10^{-17}$ s$^{-1}$ if they reach the disk \citep{umebayashiEffectsRadionuclidesIonization2009}. Taken together, CR ionization is considered as the dominant ionization source for the bulk disk due to its deep penetration. It therefore controls the coupling between gas and magnetic field, and hence regulates magnetic activities, particularly angular momentum transport in the disk. For instance, \citet{nishioFormationEarlyEvolution2025} carried out cloud core collapse simulations with simplified CR transport and demonstrated its significant impact on disk formation and early evolution. Moreover, ionization serves as a trigger for all chemical reactions in general, and hence is crucial for understanding disk chemistry \citep[e.g.,][]{obergProtoplanetaryDiskChemistry2023}.

Studies of protoplanetary disk gas dynamics and chemistry generally adopt standard prescriptions of CR ionization \citep{umebayashiFluxesEnergeticParticles1981}:
\begin{equation}
\zeta_{\rm CR}\approx\zeta_{\rm CR,0}\exp{(-\Sigma/\Sigma_{\rm CR})},
\label{eq:zetaCR}
\end{equation}
where $\zeta_{\rm CR,0}\sim10^{-17}$ s$^{-1}$ is taken for the unattenuated interstellar value \citep[e.g.,][]{spitzerHeatingRegionsEnergetic1968} and may be elevated to up to $10^{-16}$ s$^{-1}$ as inferred from molecular clouds \citep[e.g.,][]{obolentsevaReevaluationCosmicRayIonization2024}, and $\Sigma_{\rm CR}\sim96$ g cm$^{-2}$. Also note that more detailed calculations found faster-than-exponential attenuation at column densities exceeding $\Sigma_{\rm CR}$ \citep{padovaniCosmicrayIonisationCircumstellar2018}. However, such prescriptions ignore propagation effects of CRs along magnetic fields, as we elaborate below.

One line of work has emphasized modulation by stellar winds. By analogy with the heliospheric modulation of CRs by the solar wind \citep[e.g.,][]{gleesonSolarModulationGalactic1968}, \citet{cleevesEXCLUSIONCOSMICRAYS2013} showed, phenomenologically by scaling relations, that the much stronger stellar wind from T~Tauri stars could reduce the CR ionization rate incident on the disk by orders of magnitude. Chemical models further show that such reductions leave observable signatures in molecular ions and deuterated species \citep{cleevesEXCLUSIONCOSMICRAYS2014}. Another line of work highlights that CRs, being energetic charged particles, travel following the magnetic field (their gyroradii are much smaller than the disk scale). \citet{fujiiCosmicRayIonizationRate2022} showed that disk magnetic fields are dominated by the toroidal component, which enhances the path length compared to pure vertical penetration in the conventional formulation. This enhances the surface ionization rate while reducing the vertical penetration depth in a way dependent on the disk field geometry. These studies motivate treating CR ionization as a transport problem instead of prescribing an attenuated background rate.

With the establishment of MHD wind-driven disk accretion, a natural question arises. Galactic CRs must travel through the magnetized environment of the disk wind before they can reach the disk surface. It is unclear how the CR flux reaching the disk compares to the CR flux in the interstellar medium or molecular cloud. The global wind geometry, at least in ideal models, is a set of open field lines threading the disk, with field lines diverging toward infinity. Assuming CRs travel on ballistic trajectories along magnetic fields toward the disk, the CR flux should be enhanced by converging field lines, while being reduced by the magnetic mirroring effect. The scalings of these two effects with the magnetic field exactly cancel. It is thus expected that the CR ionization rate is unaffected by ballistic propagation unless the magnetic field geometry becomes more complex \citep{silsbeeMagneticMirroringFocusing2018}. This exact cancellation holds for a static field. However, particle propagation can be further subject to the advective electric field from the gas outflow, which breaks the cancellation, and more importantly, to small-scale magnetic fluctuations. Such fluctuations lead to pitch-angle scattering, parallel diffusion, field-line random walk, and nonlinear perpendicular diffusion \citep[e.g.,][]{jokipiiCosmicRayPropagationCharged1966,jokipiiStochasticAspectsMagnetic1969,matthaeusNonlinearCollisionlessPerpendicular2003,shalchiPerpendicularTransportEnergetic2020}. Further theoretical development on CR propagation through MHD disk winds is essential for a more realistic understanding of CR ionization in protoplanetary disks.

In this paper, we study test-particle transport of relativistic CR protons through a prescribed magneto-thermal protoplanetary disk wind. Using the semi-analytic wind solution of \citet{baiMAGNETOTHERMALDISKWINDS2016} as the background, we follow guiding-center trajectories along one poloidal field line originating from $1\,\mathrm{AU}$. Unresolved magnetic fluctuations are parameterized through pitch-angle-scattering and cross-field-diffusion prescriptions. We then measure the flux of CR particles reaching the wind base at the disk surface. As a first study, our calculation evaluates how prescribed pitch-angle scattering and cross-field transport filter the incident CR population through a simple 1D wind model. The aim is to identify and understand the dominant physics regulating the CR flux reaching the disk.

This paper is organized as follows. Section~\ref{sec:methods} describes the wind model, the guiding-center equations, and the prescriptions for pitch-angle scattering and cross-field diffusion. Section~\ref{sec:results} presents the
main results of our calculations of CR propagation and our interpretation.
Section~\ref{sec:discussion} discusses implications and limitations before we summarize in Section~\ref{sec:conclusions}.

\section{Model and Method}
\label{sec:methods}

We follow relativistic protons along one poloidal field line of a prescribed disk wind by integrating the guiding-center equations of motion. Pitch-angle scattering and cross-field diffusion are applied as stochastic updates after each step, with coefficients that follow from a prescribed spectrum of magnetic fluctuations. We measure the flux that reaches the wind base, normalized to the injected flux. In this section, we describe each of the model ingredients mentioned above, and all expressions are in cgs units.

\subsection{Background Wind Solution}
\label{sec:setup}

We adopt the one-dimensional magneto-thermal wind solution of \citet{baiMAGNETOTHERMALDISKWINDS2016}, in which an ideal-MHD wind is launched from the ionized surface layer of the disk. The solution is computed along a poloidal field line that corotates with its footpoint at the wind base, $(R_0,z_0)$, at the Keplerian rate $\omega=\Omega_K(R_0)$. The poloidal field line is approximated as a straight line of constant inclination $\theta$ to the midplane,
\begin{equation}
    R(s) = R_0 + s\cos\theta,
    \qquad
    z(s) = z_0 + s\sin\theta,
    \label{eq:streamline}
\end{equation}
with $s$ the arc length from the footpoint, so that $\mathrm{d}z=\mathrm{d}R\,\tan\theta$. The poloidal field strength is prescribed as
\begin{equation}
    B_p(R) \;=\; B_{p0}\,
    \frac{1+q_{\rm geo}}
         {(R/R_0) + q_{\rm geo}\,(R/R_0)^2},
    \label{eq:Bp}
\end{equation}
which interpolates between $B_p\propto R^{-1}$ near the base and $B_p\propto R^{-2}$ at large radii, with $q_{\rm geo}$ setting the transition radius. The wind solution is obtained by solving four conservation relations along the line, given the conditions at the wind base. The procedure involves matching two critical points, which determines the mass-loss rate and the Alfv\'en radius $R_A$.

We specify the poloidal field at the base by its Alfv\'en speed, $v_{Ap0}/v_K=0.1$, with $v_K$ the Keplerian speed at $R_0$. The flow is isothermal with sound speed $c_s=0.1\,v_K$, and the geometric parameters are $\theta=45^\circ$, $z_0=0.15\,R_0$, and $q_{\rm geo}=0.25$. In physical units we take a star of mass $M_*=1\,M_\odot$, $R_0=1\,\mathrm{AU}$ ($v_K=29.8\,\mathrm{km\,s^{-1}}$), and $B_{p0}=0.05\,\mathrm{G}$, which fixes the base density $\rho_0=2.2\times10^{-15}\,\mathrm{g\,cm^{-3}}$. The resulting wind-driven accretion rate is $\dot M\simeq2\times10^{-8}\,M_\odot\,\mathrm{yr^{-1}}$ (Equation~31 of \citealt{baiMAGNETOTHERMALDISKWINDS2016}), typical of Class II disks \citep{hartmannAccretionPreMainSequenceStars2016}.

\begin{figure}[t]
\centering
\includegraphics[width=\columnwidth]{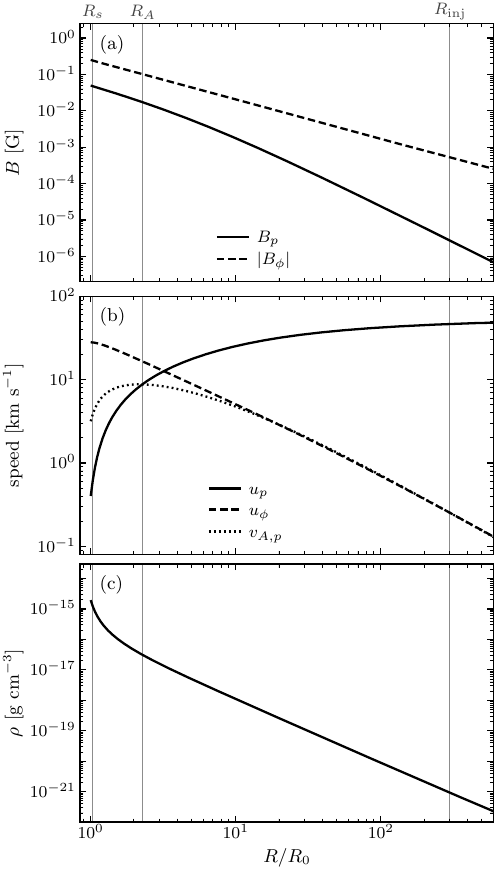}
\caption{Wind solution along the field line versus cylindrical radius $R/R_0$: (a) poloidal and toroidal field strengths $B_p$ (solid) and $|B_\phi|$ (dashed); (b) poloidal wind speed $u_p$ (solid), rotation speed $u_\phi$ (dashed), and poloidal Alfv\'en speed $v_{A,p}$ (dotted), whose base value is $v_{Ap0}$; (c) gas density $\rho$. Gray vertical lines, labeled above panel (a), mark the slow point $R_s=1.035\,R_0$, the Alfv\'en point $R_A=2.29\,R_0$, and the injection radius $R_{\rm inj}=300\,R_0$. Beyond $\sim10\,R_0$, $u_\phi$ and $v_{A,p}$ coincide.}
\label{fig:wind}
\end{figure}

Figure~\ref{fig:wind} shows the solution. The slow point lies at $R_s=1.035\,R_0$, the Alfv\'en point at $R_A=2.29\,R_0$, and the fast point at $R_f=554\,R_0$.
The toroidal field dominates everywhere: $|B_\phi|/B_p$ rises from 5.1 at the wind base
to $\approx190$ at $300\,R_0$, with $|B_\phi|\propto R^{-1}$ beyond $R_A$. The field line is therefore a tightly wound helix, and the path length along $\boldsymbol{B}$ from $300\,R_0$ to the base is $\approx4\times10^{4}\,\mathrm{AU}$. For the same reason the unit vector $\boldsymbol{b}\equiv\boldsymbol{B}/|\boldsymbol{B}|$ along the total field, $B_\phi$ included, has only a small radial component, $b_R\equiv B_R/|\boldsymbol{B}|=(B_p/|\boldsymbol{B}|)\cos\theta$, which falls from 0.14 at the base to
0.0037 at $300\,R_0$. The poloidal wind speed $u_p$ rises from $0.4\,\mathrm{km\,s^{-1}}$ at the base to $25\,\mathrm{km\,s^{-1}}$ at $10\,R_0$ and $46\,\mathrm{km\,s^{-1}}\approx1.5\,v_K$ at $300\,R_0$.
Throughout, $\boldsymbol{u}$ is the wind velocity, with poloidal and toroidal components $u_p$ and $u_\phi$, and $v$ is a particle velocity.

\subsection{Particle Injection, Boundaries, and Parameters}
\label{sec:particles}

We inject protons of kinetic energy $K_0=1\,\mathrm{GeV}$ ($\gamma=2.07$, $v=0.875\,c$) at $R_{\rm inj}=300\,R_0$, with the pitch angle $\alpha$ (not its cosine) drawn uniformly over the inward-moving hemisphere, $\pi/2<\alpha\le\pi$ ($\mu=\cos\alpha<0$, since $\boldsymbol{b}$ points outward). Sampling in $\alpha$ rather than $\mu$ places more particles near $\alpha=\pi$ and so improves the statistics in the narrow loss cone. Their gyroradius is $1.5\times10^{-6}\,\mathrm{AU}$ at the wind base and $7\times10^{-4}\,\mathrm{AU}$ at $300\,R_0$, far smaller than any scale of the wind. Therefore, we choose to use the guiding-center approximation to integrate particle trajectories. We assume particles reaching the wind base at $R_0$ are absorbed, and place the outer boundary just outside the injection radius, at $R_{\rm inj}+\delta$ (see below), which lets particles escape freely.
For each calculation, we inject $10^{6}$ particles, which is also repeated with five random seeds to help infer the statistics. We also run calculations for particles with energy $100\,\mathrm{MeV}$ and $10\,\mathrm{GeV}$, and for a different injection radius, $R_{\rm inj}=100\,R_0$, under the same setup.

In a static field, magnetic focusing and mirroring effects are expected to cancel so that the flux per unit area at the wind base equals that at the injection place \citep{silsbeeMagneticMirroringFocusing2018}. In our wind solution, the field strength rises from $B_{\rm inj}$ at $R_{\rm inj}$ to $B_b$ at the base, with $B_b/B_{\rm inj}\approx490$, so only particles moving nearly parallel to the field are inside the loss cone to avoid mirroring. Conservation of the magnetic moment sets the cosine $\mu_c$ of the cone half-angle,
\begin{equation}
    1-\mu_c^2=\frac{B_{\rm inj}}{B_b}\approx\frac{1}{490},
    \label{eq:muc}
\end{equation}
which gives a half-angle of $2.6^\circ$. The cone contains $1/490$ of the injected flux.

We define the flux ratio $\Phi$ as the flux per unit area that reaches the wind base divided by the interstellar flux per unit area, the flux that isotropic interstellar CRs carry through a static surface,
\begin{equation}
    \begin{aligned}
    \Phi
    &=
    \frac{1}{W_{\rm tot}}\,
    \sum_{i\in\mathcal{A}}
    w_i\,\frac{B(R_0)}{B(R_{\rm inj})},\\
    w_i&=\sin\alpha_i\,\max\left(|\mu_i|-\mu_{\rm w},\,0\right),
    \end{aligned}
    \label{eq:Phi}
\end{equation}
where $\mathcal{A}$ is the set of particles absorbed at the wind base $R_0$, $\alpha_i$ is the pitch angle of particle $i$ at injection, and $\mu_i=\cos\alpha_i$. The weight $w_i$ gives particle $i$ its share of the isotropic incident flux. The factor $\sin\alpha_i$ converts the uniform sampling in $\alpha$ into an isotropic distribution, and the second factor weights each direction by the rate at which it crosses $R_{\rm inj}$ inward. In a static field this rate is proportional to $|\mu_i|$. The wind, however, carries the field lines outward with $\boldsymbol{u}_\perp$, the flow velocity perpendicular to $\boldsymbol{B}$ (Section~\ref{sec:gca}). A guiding center therefore crosses $R_{\rm inj}$ inward at the radial speed $v|\mu_i|b_R-u_{\perp,R}$, which is proportional to $|\mu_i|-\mu_{\rm w}$, with $\mu_{\rm w}\equiv u_{\perp,R}/(vb_R)$ at $R_{\rm inj}$. There the field is nearly azimuthal, so $u_{\perp,R}\simeq\omega R_{\rm inj}b_R$ and $\mu_{\rm w}\simeq\omega R_{\rm inj}/v=0.034$ for the fiducial runs. Particles with $|\mu_i|<\mu_{\rm w}$ do not cross at all. The normalization $W_{\rm tot}=\sum_i\sin\alpha_i\,|\mu_i|$ runs over all injected particles. It is $\sum_iw_i$ for a static field and thus measures the interstellar flux in the same units. The factor $B(R_0)/B(R_{\rm inj})=B_b/B_{\rm inj}\approx490$ is the flux-tube weight introduced above.
A value $\Phi=1$ means no modulation by the wind, and $\Phi\to0$ corresponds to complete exclusion. 

The injection radius may be considered as the boundary between the wind and the interstellar medium (ISM; Section~\ref{sec:discussion:limitations}), although we do not model this interface explicitly. A particle that moves back out through it rejoins the ISM, and we do not follow it further: if it re-entered the wind, it would be part of the inflow that the injected flux already represents. Note that in practice we place the escape boundary slightly outward at $R_{\rm inj}+\delta$ to accommodate the cross-field diffusion operation, where we take $\delta=0.01\,R_{\rm inj}$ by default.
Varying $\delta$ between $0.003$ and $0.03\,R_{\rm inj}$ (e.g., the light bars in Figure~\ref{fig:flux_ratio}) shows that while the flux ratio $\Phi$ is modestly affected under strong turbulence, the overall effect is minor and does not affect the general results.

It is largely unknown to what level PPD winds are turbulent. While disk winds, at least those launched from the inner few AU, are expected to be largely laminar \citep{baiWINDDRIVENACCRETIONPROTOPLANETARY2013}, more recent simulations incorporating more comprehensive physics suggest the disk surface layer can be turbulent with episodic accretion flows \citep{moriRadiativeNonidealMagnetohydrodynamic2025}. We incorporate the effect of turbulence into the integration of particle trajectories (without modifying the background laminar wind solution).
The turbulence model has four parameters: amplitude $\eta\equiv(\delta B/B)^2$, spectral index $q\in\{1,\,3/2\}$, outer scale $L_c=f_cR$ with $f_c=0.1$, and a normalization $C_{\rm cap}\in\{1/6,\,1/2\}$ for cross-field diffusion (detailed in Section~\ref{sec:perp}). We scan the amplitude parameter $\eta$ over the range of $10^{-9}$--$10^{-1}$ with five points per decade (two per decade for the runs at other energies and injection radii).

We list all our physical parameters in Table~\ref{tab:params}. In the following, Section~\ref{sec:gca} gives the equations of motion, whereas Sections~\ref{sec:spectrum}--\ref{sec:perp} specify the turbulence model on particle motion and may be skipped by readers not interested in the details.

\begin{deluxetable*}{lll}
\tablecaption{Parameters of the wind, the physical units, and the particle and turbulence setup. Fiducial values are listed first, alternatives in parentheses. The critical points, the field ratio at the base, the base density, and the accretion rate are outputs of the wind solution.\label{tab:params}}
\tablehead{
    \colhead{Symbol} &
    \colhead{Description} &
    \colhead{Value}
}
\startdata
\multicolumn{3}{l}{\textit{Wind} \citep{baiMAGNETOTHERMALDISKWINDS2016}} \\
$\theta$            & Inclination of the poloidal field line                 & $45^\circ$                                   \\
$z_0/R_0$           & Height of the wind base                                & $0.15$                                       \\
$q_{\rm geo}$       & Transition parameter of $B_p(R)$, Equation~(\ref{eq:Bp}) & $0.25$                                     \\
$v_{Ap0}/v_K$       & Poloidal Alfv\'en speed at the wind base               & $0.1$                                        \\
$c_s/v_K$           & Isothermal sound speed                                 & $0.1$                                        \\
$\omega/\Omega_K$   & Angular velocity of the field line                     & $1$                                          \\
$R_A/R_0$           & Alfv\'en point                                         & $2.29$                                       \\
$|B_\phi|/B_p$      & Toroidal-to-poloidal field ratio at the wind base                           & $5.1$                                        \\
$\dot M$            & Wind-driven accretion rate                             & $2\times10^{-8}\,M_\odot\,\mathrm{yr^{-1}}$  \\
\hline
\multicolumn{3}{l}{\textit{Physical units}} \\
$M_*$               & Stellar mass                                           & $1\,M_\odot$                                 \\
$R_0$               & Footpoint radius (wind base, absorbing boundary)       & $1\,\mathrm{AU}$                             \\
$v_K$               & Keplerian speed at $R_0$                               & $29.8\,\mathrm{km\,s^{-1}}$                  \\
$B_{p0}$            & Poloidal field at the wind base                        & $0.05\,\mathrm{G}$                           \\
$\rho_0$            & Gas density at the wind base                           & $2.2\times10^{-15}\,\mathrm{g\,cm^{-3}}$     \\
\hline
\multicolumn{3}{l}{\textit{Particles and turbulence}} \\
                    & Species                                                & proton                                       \\
$K_0$               & Kinetic energy at injection                            & $1\,\mathrm{GeV}$ ($100\,\mathrm{MeV}$, $10\,\mathrm{GeV}$) \\
$R_{\rm inj}/R_0$   & Injection radius                                       & $300$ ($100$)                                \\
$\delta/R_{\rm inj}$ & Escape boundary at $R_{\rm inj}+\delta$           & $0.01$ ($0.003$--$0.03$)                \\
                    & Particles per $\eta$ point                             & $10^{6}\times5$ seeds                        \\
$\eta$              & Turbulence amplitude $(\delta B/B)^2$                  & $10^{-9}$--$10^{-1}$                         \\
$q$                 & Spectral index                                         & $3/2$ ($1$)                                  \\
$f_c$               & Outer scale, $L_c=f_c R$                               & $0.1$                                        \\
$C_{\rm cap}$       & Normalization of capped perpendicular diffusion        & $1/6$ ($1/2$)                                \\
\enddata
\end{deluxetable*}

\subsection{Guiding-Center Equations and Stochastic Updates}
\label{sec:gca}

Under the guiding-center approximation, we average over the gyration and advance particle state variables $(R,\,z,\,P_\parallel,\,\mu_{\rm mag})$, where $P_\parallel=\gamma m v_\parallel$ is the lab-frame momentum along $\boldsymbol{b}$ and $m$ is the proton mass. In our notation, quantities in the fluid (wind) frame, which moves at the fluid velocity $\boldsymbol{u}$, are denoted with a prime to distinguish them from lab-frame quantities. Since $u/c\le1.6\times10^{-4}$, we ignore higher-order relativistic corrections. The particle velocity is decomposed into $\boldsymbol{v}=v_\parallel\boldsymbol{b}+\boldsymbol{u}_\perp+\boldsymbol{v}_L$, where $\boldsymbol{u}_\perp$ is the $\boldsymbol{E}\times\boldsymbol{B}$ drift, equal to the perpendicular fluid velocity under ideal MHD, and $\boldsymbol{v}_L$ is the velocity of Larmor gyration. The magnitude of $v_L$ defines the wind-frame perpendicular momentum $p'_\perp=\gamma' mv_L$ and the magnetic moment $\mu_{\rm mag}=p'^{\,2}_\perp/(2mB)$.

The particle integration involves advancing the particle guiding-center coordinates $R$ and $z$ along the wind field trajectories following $\mathrm{d}R/\mathrm{d}t=v_\parallel b_R+u_{\perp,R}$ and $\mathrm{d}z/\mathrm{d}t=v_\parallel b_z+u_{\perp,z}$, respectively, with $v_\parallel=P_\parallel/(\gamma m)$. In a steady ideal-MHD wind, the flow is parallel to $\boldsymbol{B}$ in the frame that corotates with the field line at angular velocity $\omega$, so $\boldsymbol{u}$ is a flow along $\boldsymbol{b}$ plus the rotation $\omega R\,\hat{\boldsymbol{\phi}}$. The drift is therefore $\boldsymbol{u}_\perp=\boldsymbol{u}-(\boldsymbol{u}\cdot\boldsymbol{b})\,\boldsymbol{b}=\omega R\,(\hat{\boldsymbol{\phi}}-b_\phi\boldsymbol{b})$, with azimuthal component $\omega R\,(1-b_\phi^2)$ and poloidal components $-\omega R\,b_\phi\,(b_R,b_z)$. Its poloidal part thus lies along the field line, so $\mathrm{d}z/\mathrm{d}R=b_z/b_R=\tan\theta$, and the guiding center stays on the field line without any projection. The Runge--Kutta step described below preserves this to round-off level. Only the cross-field step of Equation~(\ref{eq:dRdiff}) moves the particle off the line, and it is mapped back. In the meantime, it remains to evolve particle parallel momentum.

The wind-frame particle parallel momentum is $p'_\parallel=\gamma' m(v_\parallel-u_\parallel)$, and the total momentum is $p'=(p'^{\,2}_\parallel+p'^{\,2}_\perp)^{1/2}=\gamma' m v'$, where $\gamma'$ and $v'$ are the wind-frame particle Lorentz factor and speed. The pitch-angle cosine in this frame is $\mu'=p'_\parallel/p'$. The lab-frame pitch-angle cosine $\mu=P_\parallel/p$, with $p^2=P_\parallel^2+2mB\mu_{\rm mag}+\gamma^2m^2u_\perp^2$, differs from $\mu'$ only at order $u/v\ll1$.

The parallel momentum obeys
\begin{equation}
\begin{aligned}
    \frac{\mathrm{d}P_{\parallel}}{\mathrm{d}t}
    =
    \gamma m\,
        \boldsymbol{u}_\perp\!\cdot\!\big(v_{\parallel}\,
        \boldsymbol{\kappa}\big)
    +\gamma m\,
        \boldsymbol{u}_\perp\!\cdot\!\big(\nabla_\parallel
        \boldsymbol{u}_\perp\big)
    -\frac{\mu_{\rm mag}}{\gamma}\,\nabla_\parallel B,
\end{aligned}
    \label{eq:gca-working}
\end{equation}
where $\boldsymbol{\kappa}=(\boldsymbol{b}\cdot\nabla)\boldsymbol{b}$ is the field-line curvature and $\nabla_\parallel=\boldsymbol{b}\cdot\nabla$. The first term couples the drift momentum $\gamma m\boldsymbol{u}_\perp$ to the curvature $\boldsymbol{\kappa}$. The second is the inertial force of the accelerating wind, and the third is the mirror force. No electric field appears explicitly, because under ideal MHD the motional electric field $-\boldsymbol{u}\times\boldsymbol{B}/c$ is already contained in the drift $\boldsymbol{u}_\perp$.
The Lorentz factor follows from the momentum components,
\begin{equation}
    \gamma
    \;=\;
    \sqrt{\,
        \frac{1 + \big(P_{\parallel}^{2}+2 m B \mu_{\rm mag}\big)/(m^{2} c^{2})}
             {1 - u_\perp^{2}/c^{2}}
    \,}.
    \label{eq:gamma}
\end{equation}
These are the standard relativistic guiding-center equations \citep{northropAdiabaticChargedparticleMotion1963,caryHamiltonianTheoryGuidingcenter2009}, written in the Galilean-invariant form of \citet{huMagnetohydrodynamicGuidingCenter2026}. In that form the curvature term contains the relative parallel velocity $v_\parallel-u_\parallel$ [Equation~(\ref{eq:app:gca-general})], but the $u_\parallel$ terms cancel against the inertial term (Appendix~\ref{app:proj:gca}), so Equation~(\ref{eq:gca-working}) contains only $v_\parallel$.

To further incorporate the effect of turbulent scattering,
we apply two stochastic updates after the deterministic step above. They are evaluated in the wind (fluid) frame, and both conserve $p'$. The first diffuses the wind-frame pitch-angle cosine,
\begin{equation}
    \Delta\mu'
    \;=\;
    \left(\frac{\partial D_{\mu'\mu'}}{\partial\mu'}\right)\!\Delta t
    + \sqrt{2\,D_{\mu'\mu'}\,\Delta t}\;\xi_{\mu'},
    \label{eq:sde-mu}
\end{equation}
which is the It\^o form of the pitch-angle Fokker--Planck equation $\partial_t f=\partial_{\mu'}(D_{\mu'\mu'}\partial_{\mu'}f)$ \citep{kloedenNumericalSolutionStochastic1992}, with $D_{\mu'\mu'}$ given in Section~\ref{sec:pitch}. The second displaces the guiding center across the field,
\begin{equation}
    \Delta R
    \;=\;
    V_{{\rm sp},R}\,\Delta t
    + \sqrt{\,2\,\kappa_\perp\,(1-b_R^{2})\,\Delta t\,}\;\xi_R,
    \label{eq:dRdiff}
\end{equation}
where the perpendicular diffusion coefficient $\kappa_\perp$ is specified in Section~\ref{sec:perp}, $V_{{\rm sp},R}$ is the It\^o drift produced by its spatial variation (Appendix~\ref{app:proj}), and $\xi_{\mu'}$ and $\xi_R$ are independent random numbers following the standard normal distribution. The factor $1-b_R^2$ is the radial projection of isotropic diffusion in the plane perpendicular to $\boldsymbol{B}$. Because $|B_\phi|\gg B_p$, one of the two perpendicular directions lies almost along the poloidal streamline, so $1-b_R^2\ge0.98$ everywhere, as in the Parker spiral of the solar wind. The other direction crosses the flux surface, and that part of a step is mapped back onto the representative field line, $R\to R+\Delta R$ and $z\to z+\Delta R\tan\theta$. This projection is the one-dimensional approximation of the model.

Equation~(\ref{eq:gca-working}) and the position advance are integrated with a fourth-order Runge--Kutta step of length $\Delta t=0.01\,B/(c\,|\mathrm{d}B/\mathrm{d}R|)$, which resolves the variation of $B$ along the line. At each stage, $\gamma$ is evaluated from Equation~(\ref{eq:gamma}). Within each step, the radial displacement of Equation~(\ref{eq:dRdiff}) is applied first and the pitch-angle update of Equation~(\ref{eq:sde-mu}) second. For numerical stability, we split the pitch-angle update into sub-steps when the estimated change is large (Appendix~\ref{app:proj}). Values of $\mu'$ outside $[-1,1]$ are reflected at the endpoints. After each update, $\mu_{\rm mag}$, $\gamma$, and $P_\parallel$ are rebuilt from the new momentum.

\subsection{Turbulence Spectrum}
\label{sec:spectrum}

We characterize (weak) turbulence in the disk wind by its power spectrum of wave intensity $I(k)$, mainly in the form of
transverse Alfv\'enic fluctuations described by a power law
\begin{equation}
    I(k) = A_q\,k^{-q},
    \qquad k \ge k_{\min},
    \label{eq:Ik}
\end{equation}
with the low-wavenumber cutoff set by the outer scale,
\begin{equation}
    k_{\min} = L_c^{-1},
    \qquad L_c = f_c\,R,
    \label{eq:Lc}
\end{equation}
and we take $f_c=0.1$. The amplitude $A_q$ is fixed by the fluctuation energy fraction
\begin{equation}
    \eta \;\equiv\; (\delta B/B)^{2}
    = \int_{k_{\min}}^{\infty} I(k)\,\mathrm{d}k ,
    \label{eq:eta}
\end{equation}
which we take to be constant along the field line.

We use $q=3/2$ as the fiducial spectrum, an Iroshnikov--Kraichnan-like scaling \citep{kraichnanInertialRangeSpectrumHydromagnetic1965}, and $q=1$ for comparison. The $q=1$ spectrum is scale-free, such that the integral in Equation~(\ref{eq:eta}) diverges. In this case, its $\eta$ is defined over a finite band $k_{\max}/k_{\min}=10^{7}$ ($\ln10^{7}=16.1$); choosing a different band would only rescale this $\eta$ by a logarithmic factor.

\subsection{Pitch-Angle Scattering and Parallel Diffusion}
\label{sec:pitch}

Resonant scattering by these fluctuations diffuses the wind-frame pitch-angle cosine. Following \citet{jaekelFokkerPlanckCoefficientsCosmic1992}, we include the finite Alfv\'en speed in the resonance condition,
\begin{equation}
    k_{\rm res} \;=\; \frac{\Omega_0}{|\mu'|\,v' + v_A},
    \label{eq:kres}
\end{equation}
where $\Omega_0=eB/(\gamma' m c)$ is the relativistic gyrofrequency, $e$ is the proton charge, and $v_A=B/\sqrt{4\pi\rho}$ is the Alfv\'en speed of the total field. The Alfv\'en speed keeps $k_{\rm res}$ finite at $\mu'=0$. Quasi-linear theory then gives \citep{jokipiiCosmicRayPropagationCharged1966,schlickeiserCosmicRayAstrophysics2002}
\begin{equation}
    D_{\mu'\mu'}
    \;=\;
    \frac{\pi}{4}\,(1-\mu'^{2})\,\Omega_0\,
    \bigl[k\,I(k)\bigr]_{k\,=\,k_{\rm res}},
    \label{eq:Dmm-res}
\end{equation}
where $kI(k)$ is the fluctuation power per logarithmic wavenumber interval. For the $q=1$ spectrum $kI(k)$ is constant over the band and
\begin{equation}
\begin{aligned}
    D_{\mu'\mu'}^{(1)}
    &=
    \varepsilon_1\,\Omega_0\,(1-\mu'^{2}),
    \\
    \varepsilon_1
    &=
    \frac{\pi\,\eta}{4\,\ln(k_{\max}/k_{\min})}.
\end{aligned}
    \label{eq:Dmm-iso}
\end{equation}
For $q>1$, Equations~(\ref{eq:Ik}) and (\ref{eq:eta}) give
\begin{equation}
    D_{\mu'\mu'}^{(q)}
    =
    \varepsilon_q\,\Omega_0\,(1-\mu'^{2})
    \left(\frac{k_{\rm res}}{k_{\min}}\right)^{1-q},
    \label{eq:Dmm-spec}
\end{equation}
with
\begin{equation}
    \varepsilon_q = \frac{\pi}{4}\,(q-1)\,\eta ,
    \label{eq:Dabs-q}
\end{equation}
so that for $q=3/2$ the coefficient falls as $(k_{\rm res}/k_{\min})^{-1/2}$. The dimensionless amplitude $\varepsilon_q$ represents the scattering rate $D_{\mu'\mu'}/[(1-\mu'^{2})\Omega_0]$ of particles resonant with the outer scale. For $q=1$, it is the same at all resonant scales. Appendix~\ref{app:proj} gives the It\^o drift terms and Appendix~\ref{app:mapping} the normalization.

As a consequence of quasi-linear pitch-angle diffusion, particles undergo spatial diffusion parallel to the mean magnetic field, with parallel diffusion coefficient following from $D_{\mu'\mu'}$ as \citep{earlDiffusiveIdealizationChargedparticle1974,schlickeiserCosmicRayAstrophysics2002}
\begin{equation}
    \kappa_\parallel
    \;=\;
    \frac{v'^{\,2}}{8}\!\int_{-1}^{1}\!
    \frac{(1-\mu'^{2})^{2}}{D_{\mu'\mu'}(\mu')}\,\mathrm{d}\mu',
    \label{eq:kpar}
\end{equation}
and the corresponding parallel mean free path is
\begin{equation}
    \lambda_\parallel
    \;\equiv\;
    \frac{3\,\kappa_\parallel}{v'}.
    \label{eq:lambdapar}
\end{equation}
For the two forms of power spectra, the integral has closed forms (Appendix~\ref{app:mapping}): $\lambda_\parallel^{(1)}=R_L/(2\varepsilon_1)$ and, for $v'\gg v_A$, $\lambda_\parallel^{(3/2)}=48\sqrt{R_L L_c}/(5\pi\eta)$, where $R_L=v'/\Omega_0$ is the gyroradius. Both scale as $1/\eta$, because $D_{\mu'\mu'}\propto\eta$. At fixed $\eta$, however, the $q=1$ spectrum puts far more power than the $q=3/2$ spectrum at the gyro-resonant scale of a $1\,\mathrm{GeV}$ proton ($k_{\rm res}/k_{\min}\ge L_c/R_L\approx4\times10^{4}$ at $R_{\rm inj}$). As a result, $\kappa_\parallel$ of the $q=1$ spectrum is smaller by a factor $\approx60$ for our fiducial model (Appendix~\ref{app:mapping}). We therefore highlight that equal values of $\eta$ do not mean equal scattering rates. We also note that the particle integration does not use $\kappa_\parallel$: parallel transport arises from the pitch-angle updates of Equation~(\ref{eq:sde-mu}). We use the closed form of $\kappa_\parallel$ only to set the capped perpendicular coefficient (Section~\ref{sec:perp}) and in diagnostics such as $\lambda_\parallel/L_c$ and $\mathcal{M}_R$ (Section~\ref{sec:results:MR}).

\subsection{Perpendicular Diffusion}
\label{sec:perp}

For weak scattering ($\lambda_\parallel\gg L_c$),
a particle
is strongly tied to the
field line. In this case, its perpendicular transport is due to the stochastic ``wandering" of field lines, known as the field-line random walk (FLRW). Streaming along the wandering line at speed $v'|\mu'|$, the particle diffuses across the mean field with a perpendicular diffusion coefficient of \citep{jokipiiStochasticAspectsMagnetic1969,matthaeusSpatialStructureFieldLine1995,shalchiPerpendicularTransportEnergetic2020}
\begin{equation}
    \kappa_{\perp,\rm FLRW}
    \;\approx\;
    \tfrac{1}{2}\,\eta\,L_c\,v'\,|\mu'|,
    \label{eq:kperp-flrw}
\end{equation}
where $\tfrac{1}{2}\eta L_c$ is taken as the diffusion coefficient of the field lines themselves.

For strong scattering, a particle no longer follows one single field line over a distance $L_c$. Nonlinear theories then give a perpendicular diffusion coefficient $a^{2}(\delta B_x^{2}/B_0^{2})\,\kappa_\parallel$, with $\delta B_x$ one transverse fluctuation component and $B_0$ the mean field \citep{matthaeusNonlinearCollisionlessPerpendicular2003,shalchiPerpendicularTransportEnergetic2020}. The two transverse components share $\eta$, so $\delta B_x^{2}/B_0^{2}=\eta/2$ and the coefficient becomes
\begin{equation}
    \kappa_{\perp,\rm cap}
    =
    C_{\rm cap}\,\eta\,\kappa_\parallel,
    \label{eq:kperp-cap}
\end{equation}
with $C_{\rm cap}=a^{2}/2$. The factor $a^{2}$ is not well understood in theory and depends on the nature of turbulence. \citet{matthaeusNonlinearCollisionlessPerpendicular2003} introduced it in their nonlinear guiding-center theory and chose $a^{2}=1/3$ to match test-particle simulations in static turbulence. That turbulence had 80\% of its energy in two-dimensional modes and 20\% in slab modes, a mix suggested by solar-wind observations \citep{bieberDominantTwodimensionalSolar1996}. With $a^{2}=1$ the expression is the fluid limit, which holds when particles diffuse along field lines that are still nearly straight. Later comparisons with simulations favor $a^{2}\approx1$ for slab and Goldreich--Sridhar turbulence and $a^{2}\approx1/3$ for the two-dimensional modes \citep[][Sections~6.5, 9.5, and 9.11]{shalchiPerpendicularTransportEnergetic2020}. \citet{shalchiPerpendicularDiffusionEnergetic2021} later reproduced these simulations without the free parameter $a^{2}$. We adopt $a^{2}=1/3$ ($C_{\rm cap}=1/6$) as the fiducial value and $a^{2}=1$ ($C_{\rm cap}=1/2$) as the upper end of the range. Therefore, empirically and for convenience, we smoothly join the two limits as a harmonic average
\begin{equation}
    \kappa_\perp^{-1}
    \;=\;
    \kappa_{\perp,\rm FLRW}^{-1}
    +
    \kappa_{\perp,\rm cap}^{-1},
    \label{eq:kperp-eff}
\end{equation}
which follows the smaller of the two. Averaging over pitch angle, the two are equal at $\lambda_\parallel=3L_c/(4C_{\rm cap})$, i.e., at $\eta\approx3\times10^{-3}$ ($C_{\rm cap}=1/6$) and $\approx10^{-2}$ ($C_{\rm cap}=1/2$) for the $q=3/2$ spectrum. Since $\kappa_\parallel\propto1/\eta$, the coefficient $\kappa_{\perp,\rm cap}$ is independent of $\eta$. Above these amplitudes $\kappa_\perp$ therefore stops growing, which is why we call $\kappa_{\perp,\rm cap}$ the cap. For a particle whose gyroradius equals the outer scale, $\kappa_{\perp,\rm cap}$ is close to the Bohm coefficient $\kappa_B=R_Lv'/3$, usually taken as the lower limit for charged-particle diffusion (Appendix~\ref{app:mapping:switch}).

Equation~(\ref{eq:kperp-flrw}) requires $\lambda_\parallel>L_c$. For a $1\,\mathrm{GeV}$ proton this holds up to the amplitude $\eta_{\rm FLRW}$ at which $\lambda_\parallel=L_c$: $\eta_{\rm FLRW}\approx2\times10^{-4}$ for the $q=1$ spectrum and $\approx1.3\times10^{-2}$ for $q=3/2$ (Appendix~\ref{app:mapping}, Figure~\ref{fig:app:lambda_lc}). Above these amplitudes the FLRW formula does not apply, but Equation~(\ref{eq:kperp-eff}) remains consistent as long as the cap takes over before $\lambda_\parallel$ falls to $L_c$. This requires $C_{\rm cap}<3/4$, which both of our values satisfy (Appendix~\ref{app:mapping:switch}). At these amplitudes, $\kappa_\perp$ is therefore set mainly by the cap.

\section{Results}
\label{sec:results}

We present the scan first: $1\,\mathrm{GeV}$ protons injected
at $R_{\rm inj}=300\,R_0$ ($R_0=1\,\mathrm{AU}$), for the $q=1$ and $q=3/2$ spectra and for
$C_{\rm cap}=1/6$ and $1/2$ (Table~\ref{tab:params}).

\subsection{The Flux Ratio versus Turbulence Amplitude}
\label{sec:results:flux}

\begin{figure*}[t]
\centering
\includegraphics[width=7.0in]{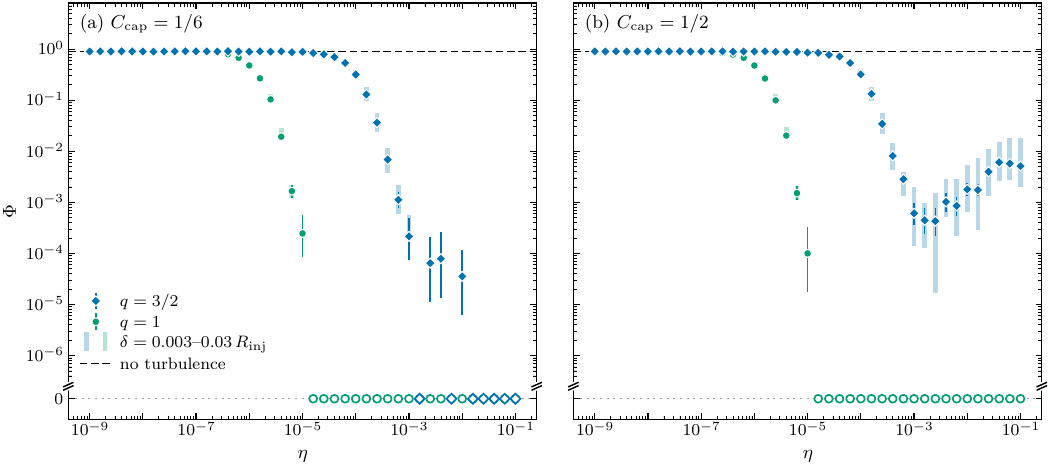}
\caption{Flux ratio $\Phi$ [Equation~(\ref{eq:Phi})] versus turbulence
amplitude $\eta$ for $1\,\mathrm{GeV}$ protons injected at
$R_{\rm inj}=300\,R_0$, for (a) $C_{\rm cap}=1/6$ (fiducial) and (b)
$C_{\rm cap}=1/2$. Blue diamonds show the $q=3/2$ spectrum and green
circles show the $q=1$ spectrum. Where the two coincide, the diamonds are drawn on
top. The dashed line is $\Phi_0=0.90$, the value without turbulence.
Error bars show $1\sigma$ Poisson errors and are mostly smaller than the symbols. Light bars show the range of $\Phi$ for $\delta=0.003$--$0.03\,R_{\rm inj}$ (Section~\ref{sec:particles}) as an approximate measure of systematic uncertainties. Open markers on the row labeled
$0$, below the axis break, mark $\Phi=0$ (no absorbed
particles). Their $1\sigma$ upper limit is $\Phi\approx2\times10^{-4}$.}
\label{fig:flux_ratio}
\end{figure*}

Figure~\ref{fig:flux_ratio}, the central result of this paper, shows the
flux ratio $\Phi$ of Equation~(\ref{eq:Phi}) against $\eta$.  Three features
stand out:
\begin{itemize}
\item A plateau at $\Phi=\Phi_0=0.90$ (slightly below 1) for $\eta\lesssim{10^{-5}}$
  ($q=3/2$) and $\eta\lesssim{2\times10^{-7}}$ ($q=1$).
\item A steep fall.  For $q=3/2$, $\Phi$ falls by three decades as $\eta$ increases from $10^{-4}$ to $10^{-3}$.  For $q=1$ the fall occurs near $\eta=10^{-6}$.
\item Strong suppression or recovery at the largest $\eta$, depending on $C_{\rm cap}$. The $q=1$ spectrum yields no particles reaching the disk for $\eta\geq1.6\times10^{-5}$ with either cap.  For the $q=3/2$ spectrum with $C_{\rm cap}=1/6$, $\Phi$ stays $\lesssim3\times10^{-4}$
for $\eta\geq10^{-3}$;
  with $C_{\rm cap}=1/2$, $\Phi$ passes through a minimum near $\eta=2\times10^{-3}$ and then rises again by about an order of magnitude.
\end{itemize}

These features can be understood qualitatively, as we describe below. Note that
the two spectra follow the same sequence, offset by a factor of about 60 in $\eta$, which is the ratio of their parallel mean free paths at fixed
$\eta$ (Section~\ref{sec:pitch}).
We defer a more quantitative understanding to the next two subsections.

First, at the low-$\eta$ limit, scattering is negligible and the particles stream freely along the field. The plateau value $\Phi_0=0.90\pm0.01$ lies slightly below unity, the value expected in a static field, where magnetic focusing and mirroring cancel exactly \citep{silsbeeMagneticMirroringFocusing2018}. Our field lines, however, turn with the rotating disk wind, and a particle moving inward along the wound-up field loses a few percent of its energy against the wind (Section~\ref{sec:results:free}). With less energy at the base it is mirrored more easily, so slightly fewer particles reach the base. The wind also carries the field lines outward, but slowly. A $1\,\mathrm{GeV}$ proton streams along the helical field from $R_{\rm inj}$ to the base in $0.8\,\mathrm{yr}$, whereas the wind needs more than $40\,\mathrm{yr}$ to cross the same radial distance. The outflow therefore reduces the CR flux entering at $R_{\rm inj}$ only slightly.

Second, as $\eta$ increases, scattering first leaves $\Phi$ almost unchanged and then suppresses it steeply.
Scattering tends to isotropize the particles, whose collective transport becomes a balance between advection by the outflow (scattering effectively occurs in the fluid frame) and diffusion along the magnetic field. As the field is tightly wound, diffusive transport has only a tiny component when projected to the radial direction. On the other hand, the flow velocity in the wind has a large radial component. Therefore, the disk outflow can substantially reduce the particle flux reaching the disk when the modulation depth $\mathcal{M}_R$ reaches order unity
(Section~\ref{sec:results:MR}).

Third, at the largest $\eta$, cross-field diffusion takes over the radial transport. Because the field is nearly toroidal, a cross-field step moves a particle almost directly in $R$, and in the case of inward diffusion, it skips the long helical path. Whether this restores part of the flux depends on the poorly constrained normalization $C_{\rm cap}$ (Section~\ref{sec:results:MR:eta}).

Given we inject $5\times10^{6}$ particles per scan point, statistics become noisy when the flux ratio drops below about $10^{-3}$, where only about ten particles reach the base, but they are sufficient to establish the strong suppression.
Moreover, we neglect collisional energy losses throughout. For $\eta\le10^{-2}$ the absorbed particles traverse at most a few ${\rm g\,cm^{-2}}$ of wind gas from $100\,\mathrm{MeV}$ to $10\,\mathrm{GeV}$ (Appendix~\ref{app:absorbed}), far below the attenuation column $\Sigma_{\rm CR}\approx10^2\,{\rm g\,cm^{-2}}$ of Equation~(\ref{eq:zetaCR}), so these losses should be minor.

\subsection{The Free-Streaming Regime}
\label{sec:results:free}

Without scattering, CRs stream freely along the field, and magnetic focusing and mirroring control the flux that reaches the base. Early studies of molecular cloud cores and collapsing clouds concluded that mirroring dominates and lowers the ionization rate by a factor of a few or more \citep{padovaniEffectsMagneticFields2011,padovaniCosmicrayIonisationCollapsing2013}. \citet{silsbeeMagneticMirroringFocusing2018} then showed with Liouville's theorem that, for an isotropic incident distribution, the two effects cancel exactly in a static field whose strength has a single maximum along the field line.
Here, we further show that considering wind motion (outflow and rotation) will lead to further CR energy loss and flux reduction.

In the laminar wind solution, $B$ rises monotonically
from $R_{\rm inj}$ to the base. The field, however,
is not static: 
the field lines corotate with the disk at the angular velocity $\omega$ of their footpoints.

In steady, axisymmetric fields, a particle conserves its energy $K+e\Phi_E$ and its canonical angular momentum $Rp_\phi+e\Psi/(2\pi c)$, where $\Phi_E$ is the electric potential, $\Psi$ the poloidal magnetic flux, and $p_\phi$ the azimuthal component of the particle momentum. In an ideal-MHD wind, each field line is an equipotential, $\Phi_E=\omega\Psi/(2\pi c)$. The particle energy in the frame that corotates with the field line, the Jacobi integral, is therefore conserved \citep{mestelAxisymmetricPulsarMagnetosphere1979,mestelAxisymmetricPulsarMagnetosphere1985}:
\begin{equation}
    \begin{aligned}
    K-\omega Rp_\phi&=\left(K+e\Phi_E\right)-\omega\left(Rp_\phi+\frac{e\Psi}{2\pi c}\right)\\
    &={\rm const}.
    \end{aligned}
    \label{eq:jacobi}
\end{equation}
Averaged over gyration, $p_\phi\simeq P_\parallel b_\phi$. The field lines trail the rotation, so a particle sliding inward along $\boldsymbol{B}$ moves in the direction of rotation ($p_\phi>0$). A particle that reaches the base lies in the narrow loss cone (Section~\ref{sec:particles}), so at injection, where $|P_\parallel|\simeq p_0$ and $|b_\phi|\simeq1$, we have $Rp_\phi\simeq R_{\rm inj}\,p_0$. At the base, we have $Rp_\phi<R_0\,p_0$, which is negligible compared to $R_{\rm inj}p_0$. Therefore, Equation~(\ref{eq:jacobi}) gives the kinetic energy at the base,
\begin{equation}
    K_b\simeq K_0-\omega R_{\rm inj}\,p_0.
    \label{eq:Kb}
\end{equation}
For a $1\,\mathrm{GeV}$ proton injected at $300\,R_0$, $K_b/K_0\approx0.95$, close to the median $K_f/K_0=0.946$ of the simulated particles that reach the base (Appendix~\ref{app:absorbed}).

The moving field lowers the flux in two ways. First, the energy loss narrows the loss cone. The magnetic moment $\mu_{\rm mag}$ is conserved, so a particle injected with pitch-angle cosine $\mu$ has $p'^{\,2}_\perp=p_0^2(1-\mu^2)\,B/B_{\rm inj}$ along its path, and it is mirrored where $P_\parallel=0$, that is, where $p'_\perp$ reaches its momentum $p$. Both relations hold to order $(u_\perp/v)^2$. Since $B$ rises and $p$ falls inward along the field line, $p'_\perp/p$ is largest at the base, and the particle reaches it if $p'_\perp\le p_b$, where $p_b$ is the momentum at $K_b$. The loss cone at injection is therefore
\begin{equation}
    1-\mu_c^2=\frac{p_b^2}{p_0^2}\,\frac{B_{\rm inj}}{B_b},
    \label{eq:losscone}
\end{equation}
which is $p_b^2/p_0^2$ times its static-field value in Equation~(\ref{eq:muc}). Second, the outflow reduces the CR flux at injection. A particle in the cone ($|\mu|\simeq1$) crosses $R_{\rm inj}$ at the radial speed $vb_R-u_{\perp,R}$, a fraction $1-\mu_{\rm w}$ of the static-field value. The cone therefore admits a fraction $(1-\mu_c^2)(1-\mu_{\rm w})$ of the interstellar flux, and each particle in it counts with the flux-tube weight $B_b/B_{\rm inj}$, so that
\begin{equation}
    \Phi_0\simeq\frac{p_b^2}{p_0^2}\,(1-\mu_{\rm w})\simeq1-\frac{3\,\omega R_{\rm inj}}{v},
    \label{eq:Phi0}
\end{equation}
where the last step assumes $\omega R_{\rm inj}\ll v$ and $|b_\phi|\simeq1$. 

One can arrive at the same results by employing Liouville's theorem. Every inward direction at the base traces back along the field to $R_{\rm inj}$, and the phase-space density $f$ is conserved along the way, so at the base $f$ is isotropic and equals its interstellar value. The flux carried within an energy interval $\mathrm{d}K$ is then proportional to $p^2f\,\mathrm{d}K$, and Equation~(\ref{eq:Kb}) maps an interval $\mathrm{d}K_0$ at injection onto $\mathrm{d}K_b=(1-\omega R_{\rm inj}/v)\,\mathrm{d}K_0$ at the base. Hence $\Phi_0=p_b^2\,\mathrm{d}K_b/(p_0^2\,\mathrm{d}K_0)$, as in Equation~(\ref{eq:Phi0}). This gives $\Phi_0\simeq{0.90}$ for the fiducial runs ($1\,\mathrm{GeV}$ protons injected at $300\,R_0$). The runs of Section~\ref{sec:results:energy} test how $\Phi_0$ depends on $v$ and $R_{\rm inj}$. For $100\,\mathrm{MeV}$ and $10\,\mathrm{GeV}$ protons at $300\,R_0$, Equation~(\ref{eq:Phi0}) gives $0.80$ and $0.91$, and for $1\,\mathrm{GeV}$ protons at $100\,R_0$ it gives $0.97$. These values match the simulated plateaus (Figure~\ref{fig:flux_ratio_sparse}).

In fact, in the absence of turbulence, this process alone can be important for CR exclusion under more extreme parameters. From Equation (\ref{eq:Phi0}), the flux ratio $\Phi_0$ in the free-streaming limit depends on the wind only through $\omega R_{\rm inj}/c=R_{\rm inj}/R_{\rm LC}$, where the light cylinder radius is
\begin{equation}
R_{\rm LC}=c/\omega\approx10^4\,\mathrm{AU}\,(R_0/1\,\mathrm{AU})^{3/2}(M_*/M_\odot)^{-1/2}.  
\end{equation}
By Equation~(\ref{eq:Kb}), the wind completely excludes CRs ($p_b\to0$) once $\omega R_{\rm inj}\ge K_0/p_0=v/(1+\gamma^{-1})$, that is, beyond $0.6\,R_{\rm LC}$ at $1\,\mathrm{GeV}$ and $0.2\,R_{\rm LC}$ at $100\,\mathrm{MeV}$ for $R_0=1\,\mathrm{AU}$.
For a wind launched at $0.1\,\mathrm{AU}$, we have $R_{\rm LC}$ as small as $\approx300\,\mathrm{AU}$, and a wind that stays laminar to $\sim100\,\mathrm{AU}$ would strongly reduce the free-streaming flux of $1\,\mathrm{GeV}$ CRs to about a quarter and completely exclude those of $100\,\mathrm{MeV}$ CRs.

\subsection{Role of Turbulent Diffusion: the Modulation Depth}
\label{sec:results:MR}

As stated earlier, for sufficiently large $\eta$, particles become more efficiently scattered toward isotropy,
and inward transport of CRs becomes a competition between radial diffusion and outward advection (by the wind). It turns out that this framework yields a good approximate description of the flux ratio measured in our calculations, as we elaborate here.

Previous studies highlighted the role of resonant interaction between CRs and MHD waves on CR penetration into molecular clouds. Such waves can be self-generated \citep{ivlevPenetrationCosmicRays2018, bambicMHDPICSimulationsCosmicRay2021} or arise from pre-existing turbulence \citep{silsbeeDiffusiveFreestreamingCosmicRay2019}, which likely lead to substantial reduction of CR flux and enhanced ionization losses. In the molecular outflow of a PPD wind, we show in Appendix~\ref{app:crsi} that ion-neutral damping is too strong to sustain CR self-generated waves for our fiducial parameters.
In addition, we find that the gas column is also small enough so that ionization loss is expected to be minor (Section~\ref{sec:results:flux}).

Under an axisymmetric field configuration, the radial diffusion coefficient of an isotropized CR population can be written as
\begin{equation}
    \kappa_{RR}
    =
    \kappa_\parallel b_R^2
    +
    \left\langle
    \kappa_{\perp}
    \right\rangle
    (1-b_R^2),
    \label{eq:kRRdiag}
\end{equation}
where $\kappa_\parallel$ is given by Equation~(\ref{eq:kpar}) and
$\kappa_\perp$ by Equation~(\ref{eq:kperp-eff}), and $\langle
X\rangle\equiv\frac{1}{2}\int_{-1}^{1}X(\mu')\,\mathrm{d}\mu'$ is the average over an isotropic pitch-angle distribution. The first term is parallel diffusion projected onto $R$. It is small because the field is tightly wound ($b_R^2\approx10^{-5}$ at $R_{\rm inj}$). The second is perpendicular diffusion, also projected onto $R$. Because the field is nearly toroidal, $1-b_R^2\approx1$ and perpendicular diffusion acts almost entirely in $R$ (Section~\ref{sec:gca}).

CR transport toward the disk is opposed by outward advection through the disk wind. Analogous to the heliospheric modulation parameter
\citep{gleesonSolarModulationGalactic1968}, we can define the radial modulation depth as the radial advection speed divided by the radial diffusion coefficient, integrated throughout the propagation radii
\begin{equation}
    \mathcal{M}_R(\eta)
    \equiv
    \int_{R_0}^{R_{\rm inj}}
    \frac{|u_R(R)|}
         {\kappa_{RR}(R;\eta)}
    \,\mathrm{d}R ,
    \label{eq:MRdiag}
\end{equation}
where $u_R=u_p\cos\theta$ is the radial component of the wind velocity. It can be considered as the ratio of the path-integrated CR diffusion timescale to the outflow advection timescale.
Diffusion dominates for $\mathcal{M}_R\ll1$, and advection dominates for $\mathcal{M}_R\gg1$. We evaluate the diffusion coefficients at the injection energy $K_0$ throughout the integral. The depth is therefore calculated from the wind profile and the prescribed turbulence, independently of the particle runs.

\begin{figure*}[t]
\centering
\includegraphics[width=7.0in]{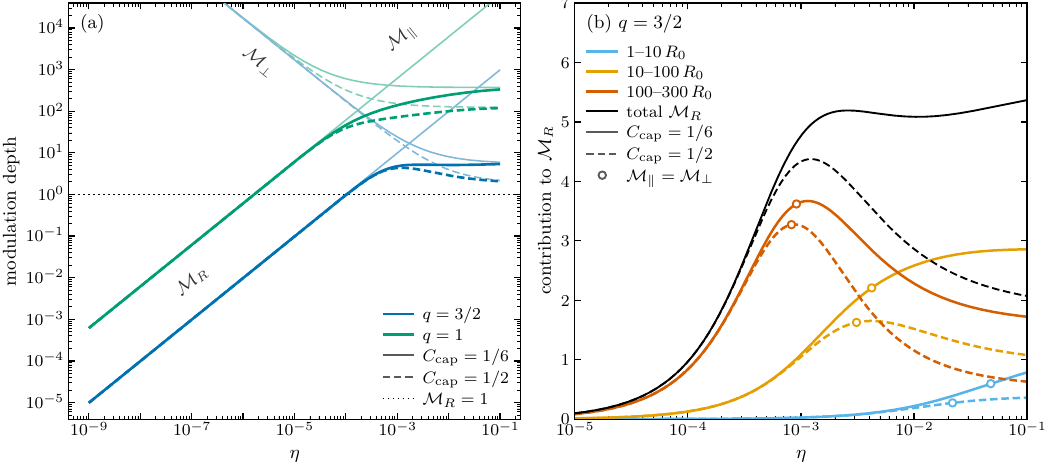}
\caption{Radial modulation depth $\mathcal{M}_R$
[Equation~(\ref{eq:MRdiag})] versus $\eta$ for the scan of Figure~\ref{fig:flux_ratio}. (a) $\mathcal{M}_R$ (thick) for
the $q=3/2$ (blue) and $q=1$
(green) spectra with $C_{\rm cap}=1/6$ (solid) and $1/2$ (dashed), and $\mathcal{M}_\parallel$ and $\mathcal{M}_\perp$, the depths from one term of $\kappa_{RR}$ alone (thin). The
dotted line marks $\mathcal{M}_R=1$. (b) Contributions of three radial decades to $\mathcal{M}_R$ (colors) and their sum (black) for $q=3/2$. Open circles mark where $\mathcal{M}_\parallel=\mathcal{M}_\perp$ within each decade.}
\label{fig:MR}
\end{figure*}

\subsubsection{Dependence on the Turbulence Amplitude}
\label{sec:results:MR:eta}

The two terms of Equation~(\ref{eq:kRRdiag}) respond oppositely to $\eta$. The parallel term falls as $1/\eta$ (Section~\ref{sec:pitch}), while the perpendicular term grows as $\eta$ on its FLRW branch and stops growing once the cap takes over (Section~\ref{sec:perp}). The depth $\mathcal{M}_R$ therefore rises with $\eta$ while the parallel term dominates $\kappa_{RR}$, and levels off or falls once the perpendicular term dominates. Figure~\ref{fig:MR}a shows both regimes, together with the depths $\mathcal{M}_\parallel$ and $\mathcal{M}_\perp$ obtained from the parallel or the perpendicular term of $\kappa_{RR}$ alone. The depth $\mathcal{M}_R$ follows $\mathcal{M}_\parallel$ at small $\eta$ and approaches $\mathcal{M}_\perp$ at large $\eta$.

At small $\eta$, the parallel term dominates and $\mathcal{M}_R\propto\eta$. It reaches unity at $\eta\approx10^{-4}$ ($\delta B/B\approx1\%$) for $q=3/2$ and at $\eta\approx1.6\times10^{-6}$ for $q=1$, where $\Phi$ has dropped to about a third of $\Phi_0$ in Figure~\ref{fig:flux_ratio}. Such weak fluctuations suffice because the radial diffusion coefficient $\kappa_{RR}$ is strongly suppressed by the $b_R^2$ factor, while their effect accumulates over the long helical path. Such fluctuations also prolong the CR travel time along the magnetic field (toward diffusive propagation), which strongly reduces the CR flux once it exceeds the outflow advection time ($\mathcal{M}_R\gtrsim1$).
Most of $\mathcal{M}_R$ builds up in the outer wind, which is fastest and most tightly wound: the wind beyond $100\,R_0$ supplies nearly 90\% of it (Figure~\ref{fig:MR}b). Asymptotically, $b_R\propto R^{-1}$ and $\kappa_\parallel\propto R$, so that $\kappa_{RR}\propto R^{-1}$ and the contribution per logarithmic interval in radius grows as $|u_R|R^2$.

At large $\eta$, the perpendicular term takes over, first in the outer wind, where $b_R$ is smallest. Because it acts almost entirely in $R$, perpendicular diffusion can lower $\mathcal{M}_R$ (whose contribution scales as $|u_R|/R$ in the integral (\ref{eq:MRdiag})), by an amount that depends on $C_{\rm cap}$. In Figure~\ref{fig:MR}b, open circles mark where the perpendicular term takes over in each radial decade ($\mathcal{M}_\parallel=\mathcal{M}_\perp$ within it). Beyond the outer circle ($\eta\approx10^{-3}$), $\kappa_\perp$ in the outer wind keeps growing until it nears the cap, so the contribution of $100$--$300\,R_0$ falls. At smaller radii the parallel term dominates until larger $\eta$, so the contributions there keep rising. For $C_{\rm cap}=1/6$ this rise roughly offsets the fall in the outer wind, and $\mathcal{M}_R$ levels off at $\approx5$. For $C_{\rm cap}=1/2$ the cap is three times higher, so $\kappa_\perp$ grows larger before it saturates. The outer contribution falls further, and the $10$--$100\,R_0$ contribution soon falls too. As a result, $\mathcal{M}_R$ drops to $\approx2$ at $\eta=0.1$, and $\Phi$ partly recovers (Figure~\ref{fig:flux_ratio}b). For $q=1$ the cap is $\approx60$ times smaller, so $\mathcal{M}_R$ stays far above unity and no particle arrives with either cap. The recovery for $C_{\rm cap}=1/2$ lies mostly above $\eta_{\rm FLRW}\approx1.3\times10^{-2}$, where the FLRW formula no longer applies (Section~\ref{sec:perp}), so it is a property of the capped model and depends on how perpendicular diffusion is treated.

\subsubsection{Flux Ratio versus Modulation Depth}
\label{sec:results:MR:phi}

\begin{figure}[t]
\centering
\includegraphics[width=\columnwidth]{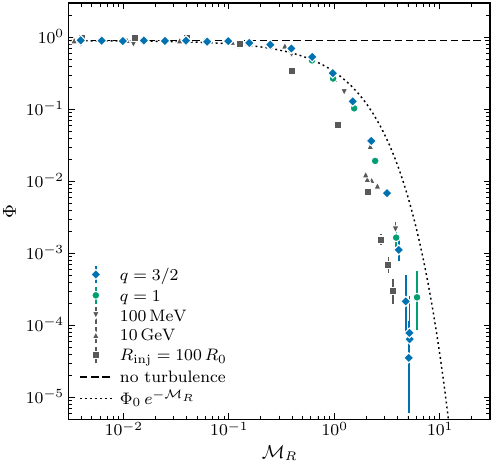}
\caption{Flux ratio $\Phi$ versus modulation depth $\mathcal{M}_R$ (error bars as in Figure~\ref{fig:flux_ratio}) for the fiducial runs with $C_{\rm cap}=1/6$ (blue diamonds, $q=3/2$; green circles, $q=1$) and the runs of Section~\ref{sec:results:energy} (gray: downward triangles, $100\,\mathrm{MeV}$; upward triangles, $10\,\mathrm{GeV}$; squares, $R_{\rm inj}=100\,R_0$), with $\mathcal{M}_R$ evaluated at each run's injection energy and radius. The dashed line is $\Phi_0=0.90$ ($1\,\mathrm{GeV}$, $300\,R_0$), and the dotted curve is $\Phi_0e^{-\mathcal{M}_R}$. Zeros and points at $\mathcal{M}_R<3\times10^{-3}$ (all at $\Phi\approx\Phi_0$) are omitted.}
\label{fig:phiMR}
\end{figure}

Figure~\ref{fig:phiMR} shows that $\Phi_0e^{-\mathcal{M}_R}$ provides an excellent estimate of the flux ratio, to within about 20\%, up to $\mathcal{M}_R\approx1$ for both spectra and all three energies at $R_{\rm inj}=300\,R_0$. The exponential is the convection--diffusion estimate of heliospheric modulation \citep{parkerPassageEnergeticCharged1965,gleesonCosmicRaysInterplanetary1967,potgieterSolarModulationCosmic2013}: where inward diffusion balances outward advection, the CR density falls inward as $\exp(-\int|u_R|\,\mathrm{d}R/\kappa_{RR})$, so the base sees the ISM density reduced by $e^{-\mathcal{M}_R}$. We can therefore use it as an empirical estimate rather than an exact law.

Beyond $\mathcal{M}_R\approx1$, $\Phi$ falls faster than the exponential, and for $R_{\rm inj}=100\,R_0$ the deviation starts at smaller $\mathcal{M}_R$. Part of the deviation likely comes from energy loss. The depth $\mathcal{M}_R$ is evaluated at $K_0$, but as scattering grows, the particles lose more energy on their way to the base. For instance, around $\mathcal{M}_R\approx2$, particle energy drops to about half of $K_0$ upon reaching the base (Appendix~\ref{app:absorbed}). A proton of lower energy has a smaller gyroradius, and hence a shorter mean free path ($\lambda_\parallel\propto\sqrt{R_L}$ for $q=3/2$ and $\propto R_L$ for $q=1$; Section~\ref{sec:pitch}), as well as a lower speed, so its $\kappa_\parallel=\lambda_\parallel v'/3$ is smaller, by about 30\% at half of $K_0$ for $q=3/2$. The actual modulation is therefore deeper than $\mathcal{M}_R$, and at $R_{\rm inj}=300\,R_0$ this can account for much of the deviation. In all cases, the flux is strongly suppressed once $\mathcal{M}_R$ exceeds unity.

\subsection{Parameter Dependence}
\label{sec:results:energy}

\begin{figure}[t]
\centering
\includegraphics[width=\columnwidth]{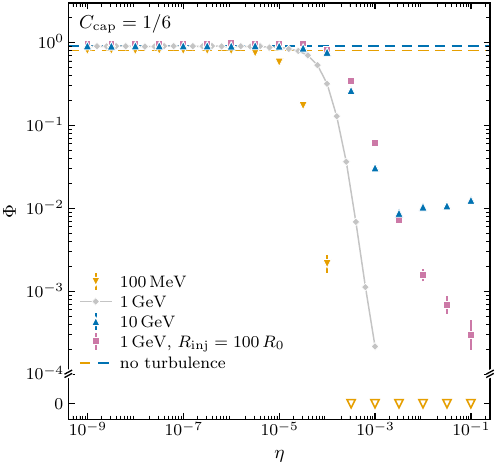}
\caption{Flux ratio $\Phi$ [Equation~(\ref{eq:Phi})] versus turbulence
amplitude $\eta$ for the $q=3/2$ spectrum at $R_{\rm inj}=300\,R_0$, with $C_{\rm cap}=1/6$.
Orange downward triangles show $100\,\mathrm{MeV}$ protons and blue
upward triangles $10\,\mathrm{GeV}$ protons. Faint gray diamonds and
curves show the fiducial $1\,\mathrm{GeV}$ results of Figure~\ref{fig:flux_ratio}a.
Purple squares show $1\,\mathrm{GeV}$ protons injected at $R_{\rm inj}=100\,R_0$ instead.
Error bars are as in Figure~\ref{fig:flux_ratio}.
Orange and blue dashed lines mark $\Phi_0$, the value without turbulence.
Open markers on the row labeled $0$, below the axis break, mark
$\Phi=0$ (no absorbed particles). Some $1\,\mathrm{GeV}$ points fall below
the plotted range.}
\label{fig:flux_ratio_sparse}
\end{figure}

Our fiducial calculations only considered $1\,\mathrm{GeV}$ particles injected from $R_{\rm inj}=300\,R_0$, with $R_0=1\,\mathrm{AU}$. Here we extend our parameters, primarily focusing on energy dependence.
Figure~\ref{fig:flux_ratio_sparse} extends Figure~\ref{fig:flux_ratio}a
to $100\,\mathrm{MeV}$ and $10\,\mathrm{GeV}$ protons at
$R_{\rm inj}=300\,R_0$, for the $q=3/2$ spectrum and $C_{\rm cap}=1/6$.
The fiducial $1\,\mathrm{GeV}$ results appear in the background for
comparison.
We see that the relation $\Phi_0e^{-\mathcal{M}_R}$ is equally applicable to CRs at lower and higher energies up to $\mathcal{M}_R\approx1$ (Figure~\ref{fig:phiMR}), affirming the robustness of this diagnostic.

Our primary finding is that lower-energy protons are more easily excluded at smaller $\eta$. As $\eta$ increases, the $100\,\mathrm{MeV}$ flux falls first and vanishes for $\eta\geq3\times10^{-4}$, while the $10\,\mathrm{GeV}$ flux falls but then recovers at $\eta>3\times10^{-3}$.
 The ordering follows from $\mathcal{M}_R$: at the same $\eta$ a lower-energy proton has a shorter mean free path, $\lambda_\parallel\propto\sqrt{R_L}/\eta$ (Section~\ref{sec:pitch}), and a lower $v'$, so $\kappa_{RR}\approx\lambda_\parallel v'\,b_R^2/3$ is smaller and $\mathcal{M}_R=1$ is reached at smaller $\eta$, by a factor of 3--4 per decade in energy. As protons below $\approx1\,\mathrm{GeV}$ are expected to produce most of the CR ionization in disks \citep{padovaniCosmicrayIonisationCircumstellar2018}, we see that they are the most sensitive to weak turbulence and are more easily excluded. On the other hand, for the more energetic $10\,\mathrm{GeV}$ CRs, $\mathcal{M}_R$ stays below 3 even at the largest $\eta$, where about 1\% of the flux still arrives thanks to more effective perpendicular diffusion.

Considering that our choice of injection location is somewhat arbitrary\footnote{In reality, it is expected to be at the analog of the heliopause, where the wind interacts with the ISM, whose location is poorly known a priori.}, we also tested our results by altering $R_{\rm inj}$. We repeated the $1\,\mathrm{GeV}$ runs with $R_{\rm inj}=100\,R_0$ (Figure~\ref{fig:flux_ratio_sparse}). The flux falls as before, but the exclusion is weaker: the fall sets in at an $\eta$ several times larger. The modulation depth explains the shift. Near the onset of exclusion, most of $\mathcal{M}_R$ in the $R_{\rm inj}=300\,R_0$ case comes from the wind beyond $100\,R_0$ (Section~\ref{sec:results:MR:eta}). Therefore, without such contributions in the $R_{\rm inj}=100\,R_0$ case, $\mathcal{M}_R$ reaches unity only near $\eta\approx10^{-3}$. Nevertheless, at a given $\mathcal{M}_R$, $\Phi$ is lower than for $R_{\rm inj}=300\,R_0$ (Figure~\ref{fig:phiMR}) by a factor of a few, which may be associated with the role of enhanced turbulence at given $\mathcal{M}_R$. At the largest $\eta$, the two injection radii give very different fluxes, which again reflects the sensitive dependence of the flux on the prescription of perpendicular diffusion (Sections~\ref{sec:results:MR:eta} and \ref{sec:discussion:limitations}).

\section{Discussion}
\label{sec:discussion}

\subsection{Implications for Cosmic-Ray Ionization in PPDs}
\label{sec:discussion:implications}

\looseness=-1 There have been limited studies of CR transport in the context of PPDs. Our work offers several new perspectives in relation to previous studies. First, we showed that the wind can exclude CRs even when it is laminar. While we confirm that magnetic mirroring and focusing effects cancel \citep{silsbeeMagneticMirroringFocusing2018}, the wind leads to additional energy loss which in the meantime narrows the loss cone (so that fewer CRs can reach the disk) and can completely exclude CRs for $R_{\rm inj}\gtrsim R_{\rm LC}$ (though this requirement is somewhat extreme). Second, we showed that if the wind is weakly turbulent, it will act similarly to CR modulation in the solar wind \citep{gleesonSolarModulationGalactic1968}, where the CR flux is set by the competition between radial diffusion and outward advection, with lower-energy CRs being preferentially excluded. This is similar to the empirical suggestion by \citet{cleevesEXCLUSIONCOSMICRAYS2013} who envisioned the exclusion to be through T~Tauri stellar winds, while we showed that the disk wind can be equally, if not more, powerful. Third, our work corroborates that of \citet{fujiiCosmicRayIonizationRate2022} who pointed out the CR ionization rate in disks should care about disk field geometry. Clearly, it must also care about the field structure in the disk wind.

Observationally, there have been limited and sometimes conflicting constraints on CR ionization rates in PPDs.
For reference, the CR ionization rate is of the order $\sim10^{-16}\,\mathrm{s^{-1}}$ in diffuse clouds \citep{indrioloInvestigatingCosmicRayIonization2012,indrioloHerschelSurveyGalactic2015,obolentsevaReevaluationCosmicRayIonization2024}.
Recently, there have been a few lines of evidence pointing to local variations of CR ionization rates \citep{pinedaProbingPhysicsStar2024,tangStarFormationDrives2026,luoHINSAViewCosmicray2026}, suggestive of locally accelerated CRs \citep[for a review, see][]{padovaniImpactLowenergyCosmic2020}.
In disks, chemical models of the HCO$^+$ and N$_2$H$^+$ emission of TW~Hya favor $\zeta_{\rm CR}\lesssim10^{-19}\,\mathrm{s^{-1}}$, which \citet{cleevesCONSTRAININGXRAYCOSMICRAY2015} took as the first sign that winds or magnetic fields exclude CRs from an accreting T~Tauri disk. In IM~Lup, the data point to a CR ionization rate that is low inside $\sim100\,\mathrm{AU}$ and high outside \citep{seifertEvidenceCosmicRayGradient2021}. Models of the MAPS disks give a midplane ionization rate of $\gtrsim10^{-18}\,\mathrm{s^{-1}}$ in IM~Lup, AS~209, and HD~163296, but $<10^{-18}\,\mathrm{s^{-1}}$ in GM~Aur \citep{aikawaMoleculesALMAPlanetforming2021}. By contrast, JWST observations of ro-vibrational H$_2$ emission from the outer disk of Oph~163131 imply an excessive rate of $\sim(1$--$10)\times10^{-15}\,\mathrm{s^{-1}}$ \citep{assaniJWSTEdgeonDisk2026}. Apart from differences in tracers and models, the spread may reflect both local CR sources and CR transport through the wind.

\looseness=-1 On one hand, we only explored limited parameters on CR transport through the MHD wind. In a future publication, we will extend our study to a broad CR spectrum, and over field lines launched over a broad range of radii toward a global picture of CR transport from the ISM toward the disk surface. In the meantime, another avenue would be to explore CR transport through local CR sources. There have been several calculations studying particle acceleration in accretion shocks~\citep{padovaniCosmicrayAccelerationYoung2015,padovaniProtostarsForgesCosmic2016,gachesExplorationCosmicrayAcceleration2018}, reconnection sites~\citep{brunnIonizationInnerTauri2023,brunnImpactsEnergeticParticles2024}, and jets~\citep{padovaniCosmicrayAccelerationYoung2015,padovaniProtostarsForgesCosmic2016}, but transport models are highly simplified concerning how these accelerated particles can propagate into the disk \citep{rabStellarEnergeticParticle2017,offnerImpactCosmicRayFeedback2019,rodgers-leeStellarCosmicRays2020}, calling for more realistic calculations \citep[e.g.,][]{fraschettiMottledProtoplanetaryDisk2018} to combine realistic models of PPD gas dynamics and CR transport.

\subsection{Limitations}
\label{sec:discussion:limitations}

As a first study, our model rests on several idealizations. We do not expect them to change the main conclusions, namely the nearly unmodulated flux in a laminar wind and the strong exclusion once $\mathcal{M}_R$ exceeds unity, but they point to directions for future improvements and applications.

The first set concerns the wind. We use a single one-dimensional magneto-thermal wind solution with fixed parameters, along one straight poloidal field line. A different field strength or inclination would change the wind profile and $\kappa_{RR}$, and hence $\mathcal{M}_R(\eta)$. The relation $\Phi\approx\Phi_0e^{-\mathcal{M}_R}$ has been tested only for this wind, and it holds best for $R_{\rm inj}=300\,R_0$. The one-dimensional model also cannot treat perpendicular diffusion self-consistently: the part of each perpendicular step that crosses flux surfaces is mapped back onto the field line (Section~\ref{sec:gca}). Neighboring flux surfaces have different field strengths, wind speeds, and paths to the disk, so following particles across them could change where and how many arrive, especially at large $\eta$. Finally, the CRs are test particles in a steady wind: the ionization they produce does not feed back on the wind.

The second set concerns the turbulence. Its amplitude, spectral index, outer scale, and the normalization $C_{\rm cap}$ are prescribed, with $\eta$ constant along the field line, and our scans cannot tell which values apply to real disk winds. The onset of exclusion lies where the FLRW formula applies and hardly depends on $C_{\rm cap}$, whereas the flux at the largest amplitudes is set by perpendicular diffusion beyond the range of that formula (Section~\ref{sec:perp}) and depends strongly on $C_{\rm cap}$. It is highly desirable for future work to address the nature of turbulence in PPD winds, which likely play a crucial role in CR transport.

The third set concerns the injection, where we set $R_{\rm inj}$ somewhat arbitrarily.
In principle, injection should take place at the interface between the wind and the ISM, whose interaction leads to the formation of a termination shock in the wind, a contact discontinuity, and a forward shock in the ISM.
In the heliosphere, the flux of $\sim100\,\mathrm{MeV}$ galactic CRs increases by a factor of several from the termination shock to the heliopause, where it reaches its interstellar level \citep{stoneVoyager1Explores2005,stoneVoyager1Observes2013,webberVoyager122014,cummingsGALACTICCOSMICRAYS2016}. Our injection radius is better understood as at the termination shock, whose location is poorly known and should depend on both the wind and the ISM, yet the CR flux is expected to be modestly reduced relative to the ISM value.
The location matters: moving $R_{\rm inj}$ from $300$ to $100\,R_0$ shifts the onset of exclusion to an $\eta$ several times larger, because the outer wind dominates $\mathcal{M}_R$ (Section~\ref{sec:results:energy}). 
A treatment of the wind--ISM interface and of the transport beyond it is left for future work.

\section{Summary}
\label{sec:conclusions}

We investigate the transport of CRs through a magnetized PPD wind. We trace test protons under the guiding-center approximation
along one prescribed poloidal field line of the magneto-thermal wind of \citet{baiMAGNETOTHERMALDISKWINDS2016}, 
with additionally prescribed pitch-angle scattering and cross-field diffusion. The fiducial runs inject $1\,\mathrm{GeV}$ protons at $R_{\rm inj}=300\,R_0$, with $R_0=1\,\mathrm{AU}$. We scan $\eta=(\delta B/B)^2$ for two spectra ($q=1$ and $3/2$) and two perpendicular diffusion normalizations ($C_{\rm cap}=1/6$ and $1/2$). The flux ratio $\Phi$ gives the particle flux at the wind base relative to the interstellar flux.
Our main results are:

\begin{enumerate}
\item In a laminar wind, the particle flux is only slightly reduced:
$\Phi_0\approx0.8$--$0.9$ for $0.1$--$10\,\mathrm{GeV}$ CRs injected at $300\,\mathrm{AU}$.
This is primarily because magnetic focusing and mirroring cancel in the static limit as found previously. The small reduction, which we newly identify here, is associated with CR energy loss as they travel against the outflow.
A particle streaming inward from $R_{\rm inj}$ reaches the base with the kinetic energy $K_b\simeq K_0-\omega R_{\rm inj}p_0$, where $\omega\approx\Omega_K(R_0)$ is the angular velocity of the field line. This leads to a flux reduction by
$\Phi_0\simeq1-3\,\omega R_{\rm inj}/v\approx1-3R_{\rm inj}/R_{\rm LC}$,
where $R_{\rm LC}=c/\omega$ is the light cylinder radius of the field line. CRs are excluded altogether once $R_{\rm inj}/R_{\rm LC}\gtrsim K_0/(p_0c)$.

\item {As scattering increases}, outward advection through the wind overtakes radial
diffusion and suppresses CR transport toward the disk. Suppression sets in when the radial modulation depth
$\mathcal{M}_R=\int_{R_0}^{R_{\rm inj}}|u_R|\,\mathrm{d}R/\kappa_{RR}$
exceeds unity. This is facilitated by the fact that the magnetic field in the wind is tightly wound, 
so only the small fraction $b_R^2$ of parallel diffusion acts radially, allowing even a slow wind to exclude CR particles.
$\Phi$ roughly follows $\Phi_0e^{-\mathcal{M}_R}$ up to $\mathcal{M}_R\approx1$ at $0.1$--$10\,\mathrm{GeV}$ and falls faster beyond. Injecting at $100\,R_0$ instead gives a lower $\Phi$ at the same $\mathcal{M}_R$.

\item At the largest $\eta$, cross-field diffusion controls the radial transport. Because the field is nearly toroidal, cross-field diffusion allows some CR particles to bypass the long helical path, and can partly restore the flux. Nevertheless, the final results depend on the details of how perpendicular diffusion is treated, and the location of the injection boundary, both being poorly constrained.

\item As turbulence amplitude increases, low-energy CRs are preferentially excluded, analogous to CR exclusion by the solar wind.
\end{enumerate}

Given our fiducial parameters, the typical threshold turbulent amplitude for CR exclusion (wind launched from $1\,\mathrm{AU}$ and for GeV CRs) is $\eta=(\delta B/B)^2\sim10^{-4}$. Therefore,  even weak magnetic fluctuations can strongly impact the CR flux reaching the inner PPDs. Such fluctuations carry little energy, but along the long, tightly wound path to the disk they are enough to isotropize the CRs, after which the outflow carries them away. A wind that is laminar on large scales can thus still keep CRs out if it has enough fluctuation power at the resonant scales, although the threshold amplitude depends on the turbulence spectrum and the wind geometry. This result points to a potential feedback loop in PPDs: CR ionization affects the gas dynamics in PPDs, whereas any perturbation in the disk would affect the level of turbulence in the wind, which strongly influences the CR flux arriving in the disk, and this in turn affects the disk gas dynamics. Currently, our model is still highly simplified in several aspects.
Future work should incorporate better treatment of the disk wind model and its interaction with the ISM, as well as improved treatment of turbulence and CR scattering in the disk wind.

\clearpage
\appendix
\renewcommand{\theHequation}{\theequation}%

\section{Guiding-Center Equations and Numerical Scheme}
\restartappendixnumbering
\label{app:proj}

\subsection{Reduction of the Parallel-Momentum Equation}
\label{app:proj:gca}

The parallel-momentum equation of
\citet{huMagnetohydrodynamicGuidingCenter2026} reads
\begin{equation}
    \frac{{\rm d}P_\parallel}{{\rm d}t}
    \;=\;
    (v_\parallel-u_\parallel)\,\boldsymbol{P}\!\cdot\!\boldsymbol{\kappa}
    + \gamma m\,\boldsymbol{u}_\perp\!\cdot\!\nabla_\parallel\boldsymbol{u}
    - \frac{\mu_{\rm mag}}{\gamma}\,\nabla_\parallel B
    + e\,E_\parallel ,
    \label{eq:app:gca-general}
\end{equation}
with the curvature vector
$\boldsymbol{\kappa}=(\boldsymbol{b}\!\cdot\!\nabla)\boldsymbol{b}$,
$\boldsymbol{\kappa}\!\cdot\!\boldsymbol{b}=0$, and
$\nabla_\parallel\equiv\boldsymbol{b}\!\cdot\!\nabla$. Ideal MHD sets
$E_\parallel=0$. The gyro-averaged momentum is
$\boldsymbol{P}=P_\parallel\boldsymbol{b}+\gamma m\,\boldsymbol{u}_\perp$,
so $\boldsymbol{P}\!\cdot\!\boldsymbol{\kappa}
=\gamma m\,\boldsymbol{u}_\perp\!\cdot\!\boldsymbol{\kappa}$. Writing
$\boldsymbol{u}=u_\parallel\boldsymbol{b}+\boldsymbol{u}_\perp$ and using
$\nabla_\parallel\boldsymbol{b}=\boldsymbol{\kappa}$ and
$\boldsymbol{u}_\perp\!\cdot\!\boldsymbol{b}=0$, the first two terms
combine to
\begin{equation}
\begin{aligned}
    &\gamma m\,\boldsymbol{u}_\perp\!\cdot\!\big[
        (v_\parallel-u_\parallel)\,\boldsymbol{\kappa}
        + u_\parallel\boldsymbol{\kappa}
        + \nabla_\parallel\boldsymbol{u}_\perp\big] \\[2pt]
    &\qquad=\;
    \gamma m\,\boldsymbol{u}_\perp\!\cdot\!\big(
        v_\parallel\boldsymbol{\kappa}
        + \nabla_\parallel\boldsymbol{u}_\perp\big),
\end{aligned}
    \label{eq:app:curv-inertia}
\end{equation}
which gives Equation~(\ref{eq:gca-working}).
We do not follow the curvature drift, a slow drift across $\boldsymbol{B}$, so the only cross-field transport is the diffusion of Section~\ref{sec:perp}.

\subsection{Projection Identities}
\label{app:proj:geom}

The field line is straight in the poloidal plane,
${\rm d}z={\rm d}R\tan\theta$ and ${\rm d}s={\rm d}R/\cos\theta$, and
the background is axisymmetric. Every background quantity $g$ is taken
to vary only along the line, $\nabla g=({\rm d}g/{\rm d}s)\,\hat{p}$
with $\hat{p}=\cos\theta\,\hat{R}+\sin\theta\,\hat{z}$, so that
\begin{equation}
    \frac{\partial g}{\partial R}
        = \cos^2\!\theta\,\frac{{\rm d}g}{{\rm d}R},
    \qquad
    \frac{\partial g}{\partial z}
        = \sin\theta\cos\theta\,\frac{{\rm d}g}{{\rm d}R}.
    \label{eq:app:projid}
\end{equation}

\subsection{Radial It\^o Drift of the Perpendicular Step}
\label{app:proj:Vsp}

The perpendicular diffusion tensor is
$\mathsf{T}=\kappa_\perp(\mathsf{I}-\boldsymbol{b}\boldsymbol{b})$, whose
$RR$ element, $T_{RR}=\kappa_\perp(1-b_R^2)$, gives the variance of the
radial step in Equation~(\ref{eq:dRdiff}). The It\^o drift is the radial
component of $\nabla\!\cdot\!\mathsf{T}$ in cylindrical coordinates.
Using Equation~(\ref{eq:app:projid}), and writing $b_\phi$ and $b_z$ for the toroidal and vertical components of $\boldsymbol{b}$, we obtain
\begin{equation}
\begin{aligned}
    V_{{\rm sp},R}
    \;=\;\;& \frac{\kappa_\perp\,(b_\phi^2-b_R^2)}{R}
    + \cos^2\!\theta\,\frac{{\rm d}}{{\rm d}R}
        \big[\kappa_\perp\,(1-b_R^2)\big] \\[2pt]
    -\;& \sin\theta\cos\theta\,\frac{{\rm d}}{{\rm d}R}
        \big[\kappa_\perp\,b_R\,b_z\big].
\end{aligned}
    \label{eq:app:Vsp}
\end{equation}
Separating the gradient of $\kappa_\perp$ from the derivatives of the
field-direction factors,
\begin{equation}
    V_{{\rm sp},R}
    \;=\;
    \frac{\kappa_\perp\,(b_\phi^2-b_R^2)}{R}
    + \mathcal{G}_R
      \left(\frac{\partial\kappa_\perp}{\partial R}\right)_{\!P_\parallel,\mu_{\rm mag}}
    + \kappa_\perp\,\mathcal{D}_R ,
    \label{eq:app:Vsp-decomp}
\end{equation}
with
\begin{equation}
\begin{aligned}
    \mathcal{G}_R
        &= \cos^2\!\theta\,(1-b_R^2)
           - \sin\theta\cos\theta\,b_R\,b_z, \\[2pt]
    \mathcal{D}_R
        &= \cos^2\!\theta\,\frac{{\rm d}(1-b_R^2)}{{\rm d}R}
           - \sin\theta\cos\theta\,\frac{{\rm d}(b_R\,b_z)}{{\rm d}R}.
\end{aligned}
    \label{eq:app:GD}
\end{equation}
Only the explicit dependence of $\kappa_\perp$ on $R$ at fixed
$(P_\parallel,\mu_{\rm mag})$ is differentiated; the dependence through
$B$ and the particle state ($\gamma'$, $\mu'$) is not. For the FLRW form,
Equation~(\ref{eq:kperp-flrw}) with $L_c=f_cR$, this gives
$(\partial\kappa_{\perp,\rm FLRW}/\partial R)_{P_\parallel,\mu_{\rm mag}}
=\kappa_{\perp,\rm FLRW}/R$. For the combined coefficient of
Equation~(\ref{eq:kperp-eff}), with $\kappa_1\equiv\kappa_{\perp,\rm FLRW}$
and $\kappa_2\equiv\kappa_{\perp,\rm cap}$,
\begin{equation}
\begin{aligned}
    \left(\frac{\partial\kappa_\perp}{\partial R}\right)_{\!P_\parallel,\mu_{\rm mag}}
    &=
    \frac{\kappa_2^{2}\,(\partial\kappa_1/\partial R)_{P_\parallel,\mu_{\rm mag}}
          + \kappa_1^{2}\,{\rm d}\kappa_2/{\rm d}R}
         {(\kappa_1+\kappa_2)^{2}}, \\[2pt]
    \frac{{\rm d}\kappa_2}{{\rm d}R}
    &= C_{\rm cap}\,\eta\,\frac{{\rm d}\kappa_\parallel}{{\rm d}R},
\end{aligned}
    \label{eq:app:dkeff}
\end{equation}
since $C_{\rm cap}$ and $\eta$ are constant along the line.

\subsection{Pitch-Angle Update}
\label{app:proj:mu}

The drift in Equation~(\ref{eq:sde-mu}),
$\partial D_{\mu'\mu'}/\partial\mu'$, is $-2\varepsilon_1\Omega_0\mu'$ for the $q=1$ spectrum,
Equation~(\ref{eq:Dmm-iso}). For $q>1$ we
substitute $k_{\rm res}=\Omega_0/h$, with $h(\mu')\equiv|\mu'|v'+v_A$,
and $k_{\min}=1/L_c$ into Equation~(\ref{eq:Dmm-spec}):
\begin{equation}
\begin{aligned}
    D_{\mu'\mu'}&=\Lambda_\eta\,(1-\mu'^{2})\,h^{\,q-1}, \\[2pt]
    \Lambda_\eta&\equiv\frac{\pi}{4}(q-1)\,\eta\,
        \Omega_0^{\,2-q}\,L_c^{\,1-q},
\end{aligned}
    \label{eq:app:Lambda}
\end{equation}
so that
\begin{equation}
\begin{aligned}
    \frac{\partial D_{\mu'\mu'}}{\partial\mu'}
    =\Lambda_\eta\Big[&-2\mu'\,h^{\,q-1} \\[2pt]
      &+(q-1)(1-\mu'^{2})\,h^{\,q-2}\,v'\,{\rm sgn}(\mu')\Big],
\end{aligned}
    \label{eq:app:drift-q}
\end{equation}
valid for $\mu'\neq0$ (the point $\mu'=0$ has measure zero).

The update divides the step into $n$ equal
sub-steps, with
\begin{equation}
\begin{aligned}
    n &= \max\!\left\{1,\;
         \left\lceil\frac{\Delta\mu'_{\rm est}}{0.4}\right\rceil\right\},
    \\[2pt]
    \Delta\mu'_{\rm est} &\equiv
    \left|\frac{\partial D_{\mu'\mu'}}{\partial\mu'}\right|\Delta t
    + 3\sqrt{2D_{\mu'\mu'}\Delta t}.
\end{aligned}
    \label{eq:app:subcycle}
\end{equation}
Here $\Delta\mu'_{\rm est}$ is the drift magnitude plus three standard deviations of the diffusive step. The value 0.4 sets when the step is split for numerical stability. In these runs, splitting is needed only at the largest $\eta$, with maximum sub-step counts of order ten. We re-evaluate $D_{\mu'\mu'}$ and its drift at each sub-step. Values outside $[-1,1]$ are reflected,
$\mu'\to2-\mu'$ for $\mu'>1$ and $\mu'\to-2-\mu'$ for $\mu'<-1$, and
then clamped to $\pm(1-10^{-10})$. The radial step,
Equation~(\ref{eq:dRdiff}), is not sub-cycled.

\section{Scattering Coefficients and the Parallel Mean Free Path}
\restartappendixnumbering
\label{app:mapping}

\subsection{Normalization of the Scattering Coefficient}
\label{app:mapping:norm}

Both spectra are normalized by the integrated amplitude
$\eta=\int_{k_{\min}}^{\infty}I(k)\,{\rm d}k$ of
Equation~(\ref{eq:eta}). For $q>1$ the integral converges, and
inserting Equation~(\ref{eq:Ik}) into the Alfv\'en-mode quasi-linear
coefficient of \citet{jaekelFokkerPlanckCoefficientsCosmic1992} gives
the amplitude $\varepsilon_q=(\pi/4)(q-1)\eta$ of
Equation~(\ref{eq:Dabs-q}). Cutting the integral at
$k_{\max}=10^{7}k_{\min}$ instead would change $\varepsilon_{3/2}$ by
$(k_{\min}/k_{\max})^{1/2}\simeq3\times10^{-4}$, which we neglect. For
$q=1$ the integral diverges logarithmically, so $\eta$ is assigned over
the band $k_{\min}\le k\le10^{7}k_{\min}$ of Section~\ref{sec:spectrum},
giving $\varepsilon_1=\pi\eta/[4\ln(k_{\max}/k_{\min})]=\pi\eta/(4\times16.1)$
in Equation~(\ref{eq:Dmm-iso}). The cutoff $k_{\max}$ affects only this normalization, not the
resonance condition.

\subsection{Parallel Mean Free Path}
\label{app:mapping:lambda}

As noted in Section~\ref{sec:pitch}, we evaluate $\kappa_\parallel$ from Equation~(\ref{eq:kpar}) and
$\lambda_\parallel=3\kappa_\parallel/v'$ in closed form only to set $\kappa_{\perp,\rm cap}$ of Equation~(\ref{eq:kperp-cap}) and for the diagnostics $\lambda_\parallel/L_c$ and $\mathcal{M}_R$. For $q=1$,
\begin{equation}
    \kappa^{(1)}_\parallel=\frac{v'^{\,2}}{6\,\varepsilon_1\,\Omega_0},
    \qquad
    \lambda^{(1)}_\parallel=\frac{R_L}{2\,\varepsilon_1},
    \label{eq:app:lambda1}
\end{equation}
with $R_L\equiv v'/\Omega_0$. For $q=3/2$ and $v'\gg v_A$,
\begin{equation}
    \kappa^{(3/2)}_\parallel=\frac{2\,v'^{\,3/2}}{5\,\Lambda_\eta},
    \qquad
    \lambda^{(3/2)}_\parallel=\frac{48}{5\pi\,\eta}\sqrt{R_L\,L_c},
    \label{eq:app:lambda32}
\end{equation}
with $\Lambda_\eta$ from Equation~(\ref{eq:app:Lambda}). The ratio of the
two mean free paths at fixed $\eta$ is
\begin{equation}
    \frac{\lambda^{(3/2)}_\parallel}{\lambda^{(1)}_\parallel}
    =\frac{24}{5\ln(k_{\max}/k_{\min})}\sqrt{\frac{L_c}{R_L}}
    =62 \quad\text{at } R_{\rm inj},
    \label{eq:app:ratio}
\end{equation}
where $R_L/L_c=2.3\times10^{-5}$. This ratio is the factor $\approx60$ of
Section~\ref{sec:pitch}.

The FLRW form of Equation~(\ref{eq:kperp-flrw}) requires $\lambda_\parallel>L_c$. Setting $\lambda_\parallel=L_c$ gives the largest amplitude at which it applies,
\begin{equation}
    \eta^{(1)}_{\rm FLRW}
    =\frac{2\ln(k_{\max}/k_{\min})}{\pi}\,\frac{R_L}{L_c},
    \qquad
    \eta^{(3/2)}_{\rm FLRW}
    =\frac{48}{5\pi}\sqrt{\frac{R_L}{L_c}} .
    \label{eq:app:etalambda}
\end{equation}
For a $1\,\mathrm{GeV}$ proton $R_L/L_c$ varies only between $1.5\times10^{-5}$ at
the base and $2.3\times10^{-5}$ at $300\,R_0$, because $B\propto R^{-1}$
in the outer wind makes $R_L\propto R\propto L_c$. Hence
$\eta_{\rm FLRW}\approx2\times10^{-4}$ for $q=1$ and
$\approx1.3\times10^{-2}$ for $q=3/2$ (Section~\ref{sec:perp}).

\begin{figure}[t]
\centering
\includegraphics[width=\columnwidth]{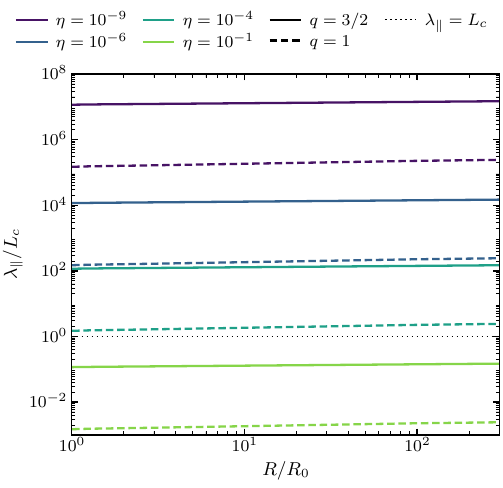}
\caption{Parallel mean free path in units of the outer scale,
$\lambda_\parallel/L_c$, along the field line for a $1\,\mathrm{GeV}$ proton at
$\eta=10^{-9}$, $10^{-6}$, $10^{-4}$, and $10^{-1}$ (color, dark to light). Solid curves show the $q=3/2$ spectrum and dashed curves show the $q=1$ spectrum. The dotted line marks $\lambda_\parallel=L_c$.}
\label{fig:app:lambda_lc}
\end{figure}

Figure~\ref{fig:app:lambda_lc} shows $\lambda_\parallel/L_c$ along the
field line for four amplitudes. At $\eta=10^{-4}$ the $q=1$ spectrum
is within a factor of two of $\lambda_\parallel=L_c$, whereas the $q=3/2$
spectrum still has $\lambda_\parallel\approx10^{2}L_c$. At
$\eta=10^{-1}$, both have $\lambda_\parallel<L_c$.

\subsection{Switch Between the Two Forms of \texorpdfstring{$\kappa_\perp$}{kappa-perp}}
\label{app:mapping:switch}

Equation~(\ref{eq:kperp-eff}) follows the smaller of $\kappa_{\perp,\rm FLRW}$
and $\kappa_{\perp,\rm cap}$. With the pitch-angle average
$\langle|\mu'|\rangle=1/2$, Equation~(\ref{eq:kperp-flrw}) gives
$\langle\kappa_{\perp,\rm FLRW}\rangle=\eta L_c v'/4$, and
Equation~(\ref{eq:kperp-cap}) with $\kappa_\parallel=\lambda_\parallel v'/3$
gives $\kappa_{\perp,\rm cap}=C_{\rm cap}\,\eta\,\lambda_\parallel v'/3$. The
two are equal at
\begin{equation}
    \lambda_\parallel=\frac{3}{4\,C_{\rm cap}}\,L_c ,
    \label{eq:app:switch}
\end{equation}
i.e., at $\lambda_\parallel=4.5\,L_c$ for $C_{\rm cap}=1/6$ and
$1.5\,L_c$ for $C_{\rm cap}=1/2$. Since $\lambda_\parallel\propto1/\eta$,
the switch occurs at $\eta_{\rm sw}=(4C_{\rm cap}/3)\,\eta_{\rm FLRW}$. This gives
$\eta_{\rm sw}\approx3\times10^{-3}$ ($C_{\rm cap}=1/6$) and
$\approx1\times10^{-2}$ ($C_{\rm cap}=1/2$) for $q=3/2$, and
$\approx4\times10^{-5}$ and $\approx1.3\times10^{-4}$ for $q=1$.

Since the FLRW form holds only below $\eta_{\rm FLRW}$ [Equation~(\ref{eq:app:etalambda})], Equation~(\ref{eq:kperp-eff}) is consistent only if the switch occurs first, that is, if $3L_c/(4C_{\rm cap})>L_c$, which requires $C_{\rm cap}<3/4$. Both of our values satisfy this condition, with $\eta_{\rm sw}\approx0.22\,\eta_{\rm FLRW}$ and $0.67\,\eta_{\rm FLRW}$. Even at $\lambda_\parallel=L_c$, the FLRW branch lowers $\kappa_\perp$ below $\kappa_{\perp,\rm cap}$ by only 18\% and 40\%, so above $\eta_{\rm FLRW}$ the cap sets the perpendicular diffusion.

The cap also behaves reasonably in strong turbulence. With $\lambda^{(3/2)}_\parallel$ from Equation~(\ref{eq:app:lambda32}), Equation~(\ref{eq:kperp-cap}) gives
\begin{equation}
    \kappa_{\perp,\rm cap}
    =\frac{16\,C_{\rm cap}}{5\pi}\,v'\sqrt{R_L L_c}
    =\frac{48\,C_{\rm cap}}{5\pi}\left(\frac{L_c}{R_L}\right)^{1/2}\kappa_B ,
    \label{eq:app:bohm}
\end{equation}
where $\kappa_B=R_Lv'/3$ is the Bohm coefficient, at which $\lambda_\parallel=R_L$. For a particle whose gyroradius equals the outer scale, $\kappa_{\perp,\rm cap}\approx0.5\,\kappa_B$ for $C_{\rm cap}=1/6$ and $\approx1.5\,\kappa_B$ for $C_{\rm cap}=1/2$, close to the Bohm value. For $1\,\mathrm{GeV}$ protons ($R_L\ll L_c$), $\kappa_{\perp,\rm cap}\approx(1$--$4)\times10^{2}\,\kappa_B$ for $q=3/2$, because this spectrum has little power at the gyro-resonant scale and so $\kappa_\parallel\gg\kappa_B$. For $q=1$, $\kappa_{\perp,\rm cap}=[2\ln(k_{\max}/k_{\min})/\pi]\,C_{\rm cap}\,\kappa_B\approx(1.7$--$5)\,\kappa_B$ at all radii, which is why $\mathcal{M}_R$ remains large for this spectrum (Section~\ref{sec:results:MR:eta}).

\section{Column Density and Energy of Absorbed Particles}
\restartappendixnumbering
\label{app:absorbed}

\begin{figure}[t]
\centering
\includegraphics[width=\columnwidth]{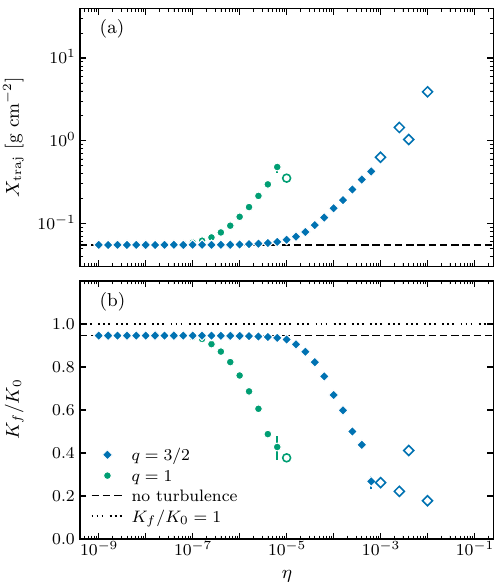}
\caption{(a) Column density $X_{\rm traj}$ and (b) kinetic-energy ratio
$K_f/K_0$ of the particles absorbed at the wind base, for
$C_{\rm cap}=1/6$, $1\,\mathrm{GeV}$, and $R_{\rm inj}=300\,R_0$. Blue diamonds ($q=3/2$) and green circles ($q=1$) show the medians over all particles absorbed in the five seeds. Error bars show $1\sigma$ bootstrap intervals and are mostly smaller than the markers. Open markers are based on fewer than five particles and have no error bars. Their scatter at the largest $\eta$, including the high point in panel (b), is due to small-number statistics. Dashed lines mark the values without turbulence, and the dotted line in panel (b) marks $K_f/K_0=1$. Note that there is no absorbed particle in the $q=1$ case for $\eta\geq1.6\times10^{-5}$, and hence no data points are shown there.}
\label{fig:absorbed_outcomes_fiducial}
\end{figure}

For every particle absorbed at the wind base we record two quantities.
The first is the column of wind gas encountered along the particle trajectory,
\begin{equation}
    X_{\rm traj}
    \;\equiv\;
    \int_{\rm traj}\rho\big(\boldsymbol{R}(t)\big)\,v(t)\,{\rm d}t ,
    \label{eq:Xtraj}
\end{equation}
where $\rho$ is the gas density of the adopted wind solution at the
guiding-center position and $v$ is the particle speed. The integral runs
from injection to arrival at the wind base and excludes the disk gas below it.
The second is the ratio $K_f/K_0$ of the kinetic energy
at absorption to that at injection. We pool the absorbed particles of the five seeds and take the median of each quantity, weighted by the injection weights $w_i$ of Equation~(\ref{eq:Phi}). Figure~\ref{fig:absorbed_outcomes_fiducial} shows these medians and their uncertainties for $C_{\rm cap}=1/6$, $1\,\mathrm{GeV}$, and $R_{\rm inj}=300\,R_0$. Note that in the $q=1$ case, there is no particle absorbed for $\eta\geq1.6\times10^{-5}$. For $q=3/2$, each point at $\eta>4\times10^{-4}$ is based on at most 13 particles from all five seeds.

The median column is about $5\times10^{-2}\,{\rm g\,cm^{-2}}$ in the non-turbulent limit. For the $q=3/2$ case, it rapidly increases with $\eta$ from $\eta\gtrsim10^{-5}$, reaching about $0.5\,{\rm g\,cm^{-2}}$ near $\eta=10^{-3}$ and at most a few ${\rm g\,cm^{-2}}$ at larger $\eta$. It grows with $\eta$ because absorbed particles spend more time in the wind before reaching the base.
Similarly, the energy lost to the motional electric field of the outflow grows with the same residence time. For the $q=3/2$ spectrum, $K_f/K_0$ falls from $0.946$ without turbulence to about 0.25 near $\eta=10^{-3}$. The value without turbulence is close to $K_b/K_0$ (Section~\ref{sec:results:free}). The $q=1$ spectrum follows the same trends in both cases, with the transition at $\approx60$ times smaller $\eta$ compared to the $q=3/2$ case. The median columns at $100\,\mathrm{MeV}$ and $10\,\mathrm{GeV}$ also stay below a few ${\rm g\,cm^{-2}}$ for $\eta\le10^{-2}$.

\section{Ion-Neutral Damping of CR Self-Generated Waves}
\restartappendixnumbering
\label{app:crsi}

\begin{deluxetable}{cccccc}
\tablecaption{Growth condition of CR self-generated waves along the wind, for a CR energy density of $1\,\mathrm{eV\,cm^{-3}}$ and $x_i=10^{-4}$, with $V_D$ from Equation~(\ref{eq:app:VD}). The waves can grow only where $V_D>V_{\rm crit}$, i.e., $\Gamma_{\max,0}/\Gamma_d>1$.\label{tab:app:crsi}}
\tablehead{
    \colhead{$R/R_0$} &
    \colhead{$\zeta$} &
    \colhead{$v_{A,i}$ ($\mathrm{km\,s^{-1}}$)} &
    \colhead{$V_{\rm crit}/c$} &
    \colhead{$V_D/c$} &
    \colhead{$\Gamma_{\max,0}/\Gamma_d$}
}
\startdata
2.3 & $2.3\times10^{-8}$  & 2290 & $3.3\times10^{5}$ & 0.080  & $2.2\times10^{-7}$ \\
10  & $3.7\times10^{-6}$  & 2440 & $2.2\times10^{3}$ & 0.017  & $4.0\times10^{-6}$ \\
100 & $4.6\times10^{-3}$  & 2210 & 1.6               & 0.011  & $2.4\times10^{-3}$ \\
200 & 0.039               & 2160 & 0.19              & 0.021  & 0.073 \\
300 & 0.13                & 2140 & 0.062             & 0.030  & 0.42 \\
\enddata
\end{deluxetable}

Here we check whether CRs can excite their own waves inside $R_{\rm inj}$ (Section~\ref{sec:results:MR}), following \citet{plotnikovInfluenceIonNeutralDamping2021}. The whole calculation assumes a purely laminar wind, as in our derivation of $\Phi_0$ (Section~\ref{sec:results:free}), which represents the maximum CR flux (and hence streaming speed) through the wind. We take a CR energy density of $1\,\mathrm{eV\,cm^{-3}}$, i.e., $n_{\rm CR}\approx10^{-9}\,\mathrm{cm^{-3}}$ for CRs of $\sim1\,\mathrm{GeV}$, and an ionization fraction $x_i=n_i/n_n=10^{-4}$. This ionization fraction is comparable to the gas-phase carbon abundance and is typical of the disk atmosphere, where FUV radiation ionizes carbon while the bulk gas remains molecular \citep{perez-beckerSURFACELAYERACCRETION2011}. We therefore take C$^+$ ions of mass $m_i=12\,m_p$ and H$_2$-dominated neutrals of mean mass $m_n=2.3\,m_p$, with a momentum transfer rate coefficient $\langle\sigma v\rangle=1.9\times10^{-9}\,\mathrm{cm^3\,s^{-1}}$ for their collisions. The field strength $B$ and the density $\rho\approx m_nn_n$ come from the wind solution (Figure~\ref{fig:wind}).

CRs that drift along $\boldsymbol{B}$ relative to the gas at a mean speed $V_D$ excite resonant Alfv\'en waves. Without damping, the fastest-growing waves grow at the rate
\begin{equation}
    \Gamma_{\max,0}\approx0.28\,\Omega_i\,\frac{n_{\rm CR}}{n_i}\left(\frac{V_D}{v_{A,i}}-1\right),
    \label{eq:app:Gmax}
\end{equation}
where $\Omega_i=eB/(m_ic)$ is the ion gyrofrequency and $v_{A,i}=B/\sqrt{4\pi n_im_i}$ is the ion Alfv\'en speed. Ion-neutral collisions damp the waves at $\Gamma_d\approx\nu_{in}/2$, where $\nu_{in}=\rho_n\langle\sigma v\rangle/(m_i+m_n)$ is the ion-neutral collision frequency. This damping rate applies when the neutrals carry most of the mass and the waves oscillate faster than $\nu_{in}$. Both conditions hold inside $R_{\rm inj}$: $\rho_i/\rho_n=5.2\times10^{-4}$, and the fastest-growing waves have frequencies above $10\,\nu_{in}$. The waves grow only if $\Gamma_{\max,0}>\Gamma_d$, that is, if
\begin{equation}
    V_D>V_{\rm crit}\equiv v_{A,i}\left(1+\frac{1}{\zeta}\right),
    \label{eq:app:Vcrit}
\end{equation}
where
\begin{equation}
    \zeta=0.56\,\frac{\Omega_i\,n_{\rm CR}}{n_i\,\nu_{in}}.
    \label{eq:app:zeta}
\end{equation}

To estimate $V_D$, we work in the frame that corotates with the field line at $\omega$. In this frame, the gas velocity $\boldsymbol{u}-\omega R\,\hat{\boldsymbol{\phi}}$ is parallel to $\boldsymbol{B}$ (Section~\ref{sec:gca}), so the ideal-MHD electric field vanishes. This velocity has the same poloidal component $u_p$ as $\boldsymbol{u}$, so the gas moves outward along the field line at $u_p/b_p$, where $b_p=B_p/B$. This relative speed is $\approx2\,\mathrm{km\,s^{-1}}$ at the base, where the gas nearly corotates with the field line, but approaches $\omega R$ ($8.9\times10^3\,\mathrm{km\,s^{-1}}$ at $R_{\rm inj}$) far beyond $R_A$, where the gas hardly rotates and the field line sweeps past it (Figure~\ref{fig:wind}b).

The CR guiding centers have no electric drift in this frame, so they move only along the field line. Since the field is also static in this frame, mirroring conserves the CR energy, and most CRs are mirrored back before they reach the base. In a steady state, the net number of CRs passing through a cross section of a flux tube per unit time is therefore the same all along the tube and equals the rate at which the base absorbs them. By the definition of $\Phi$ [Equation~(\ref{eq:Phi})], the absorbed flux per unit area at the base is $\Phi_0$ times the interstellar flux per unit area through a static surface, $n_{\rm ISM}\,v/4$, where $n_{\rm ISM}$ is the interstellar CR density. The cross section of the flux tube scales as $1/B$, so the net flux per unit area elsewhere is $\Phi_0\,n_{\rm ISM}\,v\,B/(4B_b)$. Away from the base, the CR density stays close to $n_{\rm ISM}$ (Section~\ref{sec:results:free}), so in this frame the CRs move inward along the field at a mean speed $\Phi_0\,v\,B/(4B_b)$, about $120\,\mathrm{km\,s^{-1}}$ at $R_{\rm inj}$. Mirroring thus keeps the CR inflow small, but it ties the CRs to this frame rather than to the gas, which slides past them at $u_p/b_p$. Because the gas moves outward and the CRs inward, the two speeds add up:
\begin{equation}
    V_D\approx\frac{u_p}{b_p}+\frac{\Phi_0\,v\,B}{4B_b},
    \label{eq:app:VD}
\end{equation}
with $v=0.875\,c$ and $\Phi_0=0.90$ from the fiducial runs.

Table~\ref{tab:app:crsi} lists the results. Since $\zeta\propto B/(x_in_n^2)$ and $n_n$ falls steeply outward, $\zeta$ rises from $10^{-11}$ at the base to $0.13$ at $R_{\rm inj}$. Because $v_{A,i}$ varies little, the critical drift $V_{\rm crit}$ falls outward, drops below $c$ only beyond $\approx120\,R_0$, and reaches $1.9\times10^4\,\mathrm{km\,s^{-1}}$ at $R_{\rm inj}$. Inside $R_{\rm inj}$, the actual drift $V_D$ stays below $V_{\rm crit}$ and comes closest to it at $R_{\rm inj}$, where $V_{\rm crit}/V_D=2.07$ and $\Gamma_{\max,0}/\Gamma_d=0.42$ for our fiducial model. Ion-neutral damping therefore dominates throughout this region, by more than a factor of 2 even at $R_{\rm inj}$. On the other hand, given the modest margin and the approximate nature of this estimate, self-generated waves may still become relevant for a higher CR energy density, or if the wind can propagate further out.

\clearpage

\bibliography{main}{}
\bibliographystyle{aasjournalv7}
\end{document}